\documentclass[11pt]{article}
\pdfoutput=1
\usepackage[english]{babel}
\usepackage[utf8]{inputenc}
\usepackage{graphicx}
\usepackage{array}
\usepackage{booktabs}
\usepackage{cite}
\usepackage{epsfig}
\usepackage{amsmath}
\usepackage{amssymb}
\usepackage{latexsym,url}
\usepackage{hyperref}
\usepackage{color}
\usepackage{mathtools}
\usepackage{mcite}

\def\jhep #1#2#3 {{JHEP} {\bf#1} (#2) #3}
\def\plb #1 #2 #3 {{Phys.~Lett.} {\bf B#1} (#2) #3}
\def\npb #1 #2 #3 {{Nucl.~Phys.} {\bf B#1} (#2) #3}
\def\epjc #1#2#3 {{Eur.~Phys.~J.} {\bf C#1} (#2) #3}
\def\epjd #1#2#3 {{Eur.~Phys.~J.} {\bf D#1} (#2) #3}
\def\zpc #1 #2 #3 {{Z.~Phys.} {\bf C#1} (#2) #3}
\def\jpg #1 #2 #3 {{J.~Phys.} {\bf G#1} (#2) #3}
\def\prd #1#2#3 {{Phys.~Rev.} {\bf D#1} (#2) #3}
\def\prep #1 #2 #3 {{Phys.~Rep.} {\bf#1} (#2) #3}
\def\prl #1 #2 #3 {{Phys.~Rev.~Lett.} {\bf#1} (#2) #3}
\def\mpl #1 #2 #3 {{Mod.~Phys.~Lett.} {\bf#1} (#2) #3}
\def\rmp #1 #2 #3 {{Rev. Mod. Phys.} {\bf#1} (#2) #3}
\def\cpc #1 #2 #3 {{Comp. Phys. Commun.} {\bf#1} (#2) #3}
\def\sjnp #1 #2 #3 {{Sov. J. Nucl. Phys.} {\bf#1} (#2) #3}
\def\xx #1 #2 #3 {{\bf#1}, (#2) #3}
\def\hepph #1 {{\tt hep-ph/#1}}

\newcommand{\be}{\begin{equation}}

\newcommand{\ee}{\end{equation}}
\newcommand{\bea}{\begin{eqnarray}}
\newcommand{\eea}{\end{eqnarray}}

\newcommand{\smallh}{{\scriptscriptstyle H}}
\newcommand{\mh}{m_\smallh}
\newcommand{\mt}{m_t}
\newcommand{\mb}{m_b}

\newcommand{\mathampl}{\mathcal{M}}
\newcommand{\pth}{p_{\bot}^H}

\newcommand{\divtext}{\text{div}}
\newcommand{\regtext}{\text{reg}}

\newcommand{\powheg}{{\tt POWHEG}}
\newcommand{\powhegbox}{{\tt POWHEG-BOX}}
\newcommand{\mcatnlo}{{\tt MC@NLO}}
\newcommand{\hres}{{\tt HRes}}
\newcommand{\hqt}{{\tt HqT}}

\begin{document}

\newenvironment{appendletterA}
 {
  \typeout{ Starting Appendix \thesection }
  \setcounter{section}{0}
  \setcounter{equation}{0}
  \renewcommand{\theequation}{A\arabic{equation}}
 }{
  \typeout{Appendix done}
 }

\begin{titlepage}
\nopagebreak

{\flushright{
        \begin{minipage}{5cm}
         DESY 15-067 \\
         TIF-UNIMI-2015-3 \\ 
        \end{minipage}        }}

\renewcommand{\thefootnote}{\fnsymbol{footnote}}
\vskip 1.5cm
\begin{center}
\boldmath
{\Large\bf The Higgs transverse momentum distribution in gluon fusion}\\[9pt]
{\Large\bf as a multiscale problem}\\[9pt]
\unboldmath
\vskip 1.cm

{\large E.~Bagnaschi\footnote{Email: emanuele.bagnaschi@desy.de}}

\vskip .2cm
{\it DESY, Notkestraße 85, D–22607 Hamburg, Germany
} \\[5mm]

{\large A.~Vicini\footnote{Email: alessandro.vicini@mi.infn.it}}

\vskip .2cm
{\it Tif lab, Dipartimento di Fisica, Universit\`a degli Studi di Milano and
INFN, Sezione di Milano,\\
Via Celoria 16, 
I-20133 Milano, Italy} \\[1.5mm]
\end{center}

\begin{abstract}
We consider Higgs production in gluon fusion and in particular the prediction of the Higgs transverse momentum distribution.
We discuss the ambiguities affecting the matching procedure between 
fixed order matrix elements and the resummation to all orders
of the terms enhanced by $\log(p_T^H/m_H)$ factors.
Following a recent proposal \cite{Grazzini:2013mca}, we argue that the gluon fusion process, computed considering two active quark flavors, is a 
multiscale problem from the point of view of the resummation of the collinear singular terms.
We perform an analysis at parton level of the collinear
behavior of the $\mathcal{O}(\alpha_s)$ real emission amplitudes;
relying on the collinear singularities structure of the latter,
we derive an upper limit to the range of transverse
momenta where the collinear approximation is valid.
This scale is then used as the value of the resummation scale in the analytic resummation framework
or as the value of the $h$ parameter in the \powhegbox~code.
A variation of this scale can be used to generate an uncertainty band associated to the matching
procedure.
Finally, we provide a phenomenological analysis in the Standard Model, in the Two Higgs Doublet Model and in the Minimal Supersymmetric Standard Model.
In the two latter cases, 
we provide an {\it ansatz} for the central value of the matching parameters
not only for a Standard Model-like Higgs boson, but also for heavy scalars and in scenarios where the bottom quark may play the dominant role.
\end{abstract}


\vfill
\end{titlepage}    

\setcounter{footnote}{0}
\tableofcontents

\clearpage
\section{Introduction}
 A new state with a mass of approximately 125 GeV has been observed at the LHC~\cite{Aad:2012tfa,Chatrchyan:2012ufa}.
Many investigations are under way to determine its properties
and to test its compatibility with the Higgs scalar boson of the
Standard Model (SM). 
The precise measurements of the total production cross section 
and of the branching ratios in the different allowed decay channels~\cite{Dittmaier:2011ti,Dittmaier:2012vm}
have shown that the new state couples to the known fermions and gauge bosons following the SM predictions.
Other studies target the kinematics of the decay products
to distinguish among the various spin-parity combinations~\cite{Heinemeyer:2013tqa}. 
Finally, further work will be necessary 
to clarify the structure of the scalar potential.

In the SM the main production mode of the Higgs boson at hadron colliders 
is through the gluon fusion mechanism.
The coupling of the Higgs boson to the gluons is mediated by a loop of colored particles,
with the largest contribution to the process given by the top quark.
The gluon fusion cross section is very well approximated 
by a Heavy Quark Effective Field Theory (HQEFT), 
where the Higgs mass is considered very small with respect to the one of the top quark. 
The coupling of the Higgs boson to gluons is then proportional to the Fermi constant and to the strong coupling, but it is independent of the top Yukawa coupling.
In this approach the very large NLO and NNLO-QCD corrections to the LO process 
($+100\%$ and $+30\%$ of the LO result respectively)
have been evaluated in refs.~\cite{Dawson:1990zj,Djouadi:1991tka,Spira:1995rr} and in refs.~\cite{Harlander:2000mg,Harlander:2001is,Catani:2001ic,Harlander:2002wh,Anastasiou:2002yz,Ravindran:2003um}. Recently, expressions for the N$^3$LO 
corrections have been published in refs.~\cite{Moch:2005ky,Ravindran:2006cg,Ball:2013bra,Buehler:2013fha,Anastasiou:2014vaa,Anastasiou:2015ema}.
The calculation using the complete SM Lagrangian was done up to NLO-QCD~\cite{Harlander:2005rq,Anastasiou:2006hc,Aglietti:2006tp,Bonciani:2007ex}. 
The exact treatment of the quark loops (mostly from top, bottom and charm) 
at NLO-QCD 
yields an $\mathcal{O}(-1\%)$ correction, for an Higgs with a mass of $m_H \simeq 125$ GeV and a collision energy $\sqrt{S}=14$ TeV.
Moreover finite top-mass effects at NNLO-QCD have been estimated and found to be of $\mathcal{O}(1\%)$~\cite{Marzani:2008az,Harlander:2009bw,Harlander:2009mq,Pak:2009bx,Pak:2009dg,Harlander:2009my}.
Beyond fixed-order QCD corrections, also soft-gluon resummation effects are available~\cite{Kramer:1996iq,Catani:2003zt,Idilbi:2005ni,Idilbi:2006dg,Ahrens:2008nc}. 
Moreover, the first-order electroweak (EW) contributions have been evaluated in refs.~\cite{Djouadi:1994ge,Djouadi:1997rj,Aglietti:2004nj,Aglietti:2004ki,Degrassi:2004mx,Actis:2008ug,Actis:2008ts,Bonciani:2010ms} and an estimate of the mixed 
QCD-EW contributions has been presented in ref.~\cite{Anastasiou:2008tj}. 
The PDF and $\alpha_s$ uncertainties on the total Higgs production cross section have been studied in ref.~\cite{Demartin:2010er}.

The production cross section of a Higgs boson at large transverse momentum has been computed at LO-QCD,
retaining the full quark-mass dependence, in refs.~\cite{Ellis:1987xu,Baur:1989cm}.
The NLO-EW corrections to this observable have been considered in refs.~\cite{Keung:2009bs,Brein:2010xj} in the HQEFT limit.
The NLO-QCD corrections, in the HQEFT, have been computed~\cite{deFlorian:1999zd,Ravindran:2002dc,Glosser:2002gm}.
An estimation of top-mass effects at NLO-QCD has been presented in ref.~\cite{Harlander:2012hf}.
The first results towards the determination of the Higgs production at large transverse momentum,
in the HQEFT, with NNLO-QCD accuracy, have been presented in ref.~\cite{Boughezal:2013uia}.

In this paper we want to reconsider the uncertainties
that affect the theoretical prediction for the Higgs boson transverse momentum $p_{\bot}^H$.
The transverse momentum distribution is an observable generated by QCD radiation. 
In the region of small $p_{\bot}^H$ the presence of terms enhanced by large $\log(p_{\bot}^H/m_H)$ factors 
spoils the accuracy of the fixed-order results;
in order to obtain a physically meaningful prediction these logarithms have to be resummed. 
Various techniques are available to perform the resummation. Once the latter is achieved, the resummed result 
has to be matched to the fixed-order one. Particular care is required to avoid the double counting of those logarithmic contributions
that are present in both computations. The matching procedure introduces additional unphysical variables, the matching parameters, that
define how the spectrum is divided into a soft region, where the resummed result is indeed applied, and a hard region where the fixed-order result is instead considered
as the correct description of the spectrum. 

In the HQEFT framework, the corrections up to NLO-QCD for the Higgs transverse momentum distribution
have been analytically computed and matched with the transverse momentum resummation at NNLL accuracy. 
The results have been originally implemented in the code \hqt\cite{Bozzi:2003jy,Bozzi:2005wk,Bozzi:2007pn} 
and later in the parton Monte Carlo program \hres\cite{deFlorian:2012mx}.
A similar discussion, in the Soft Collinear Effective Theory (SCET) approach, has been presented in ref.~\cite{Chiu:2012ir,Becher:2012yn,Neill:2015roa}.
In the context of matched NLO+Parton Shower (PS) Monte Carlo event generators, 
which implement the resummation algorithmically in the computer code, 
the results in the HQEFT, for Higgs production via gluon fusion,
have been presented in refs.~\cite{Frixione:2002ik,Alioli:2008tz}.
Two shower Monte Carlo codes that retain the NNLO-QCD accuracy on the inclusive observables, in the HQEFT, 
have been presented in refs.~\cite{Hamilton:2013fea,Hoche:2014dla}.

Despite the fact that the exact matrix elements retaining the full dependence on the quark masses 
were available for quite some time, they have been implemented in a NLO+PS Monte Carlo 
for the first time in ref.~\cite{Bagnaschi:2011tu}, in the \powheg~approach, and later in \mcatnlo\cite{frixione-masses}.
A similar study, in the framework of analytic resummation,
has been presented in ref.~\cite{Mantler:2012bj} and later in ref.~\cite{Grazzini:2013mca}.
Recently, these effects have been implemented in the {\tt NNLOPS} code~\cite{Hamilton:2015nsa}.
Quark mass effects have also been discussed for observables like the jet veto distribution in refs.~\cite{Banfi:2013eda,Neumann:2014nha}.
Moreover, in ref.~\cite{Banfi:2013eda}, the structure of the collinear singularities and
of the regular terms present in gluon fusion at $\mathcal{O}(\alpha_s)$ is analyzed in detail.

In ref.~\cite{Grazzini:2013mca} it has been pointed out that the matched computation of the Higgs transverse momentum distribution is a problem with three scales, namely the Higgs mass, the internal quark mass and the transverse momentum of the Higgs boson. The matching prescription between fixed-order and resummed results should account for all these scales, to avoid, as far as possible, the inclusion of spuriously large higher-order terms in the final result.
It should be noted that the presence of non-negligible interference effects between the top and the bottom quarks
assigns a simultaneous active role to both internal quarks present in the scattering amplitude.

In the framework of SCET, the separation between the singular regions where a resummation is needed and the corresponding regular parts 
has been discussed in refs. \cite{Ligeti:2008ac,Abbate:2010xh} 
with the introduction of appropriate profile functions at the level of the hadronic cross section;
this approach has been applied to Higgs studies in ref. \cite{Berger:2010xi}.
The problem of the determination of a sensible value for the scale that separates the two transverse momentum regions, the one where the resummation is needed and the one where a fixed-order description is reliable, has been discussed in QCD, at the level of the partonic cross section, in ref.~\cite{cerntalk}. 
Recently, in ref.~\cite{Harlander:2014uea}, the determination of these scales has been realized in QCD,
with an approach that exploits some general properties of the Higgs transverse momentum distribution at hadron level, 
to derive the largest interval of transverse momenta where the resummed expression can be applied.


In this paper we elaborate the approach of ref.~\cite{cerntalk}, and present a derivation at parton level
of the interval of transverse momenta where the collinear approximation of the squared matrix element is accurate and the transverse momentum resummation can be safely applied.
A comparison of the present results against those of refs.~\cite{Harlander:2014uea,Mantler:2015vba} is currently ongoing 
\cite{Bagnaschi:2015bop}.

Higgs production via gluon fusion may provide interesting information
about possible signals of physics beyond the Standard Model (BSM),
like those possible in the Two Higgs Doublets Model (2HDM) or
those predicted in the Minimal Supersymmetric Standard Model (MSSM),
thanks to the possible exchange in the loop of new colored particle that act as mediators of the interaction between the gluons and the Higgs boson.
The total cross section for Higgs production
(see ref.~\cite{Bagnaschi:2014zla} for a recent review) 
and the Higgs transverse momentum distributions 
provide complementary information 
(see e.g. ref.~\cite{Bagnaschi:2011tu})
to disentangle the SM from MSSM.
The possibility of extracting sensible information from the data
depends on the accuracy of the prediction of the $\pth$ distribution,
and, among others, on the choice of the matching parameters.

The outline of the paper is the following. 
In section~\ref{sec:generalresumm} we recall the basic elements
needed to formulate the transverse momentum resummation 
and to match the corresponding expression with fixed order results;
we make some comments on the analytic procedure and discuss in more detail the NLO+PS Monte Carlo formulation.
In particular we discuss in both cases the role of the  scales associated to the matching and we describe the differences between the two approaches.
In section~\ref{sec:hggcoll} we discuss in detail the $gg \to gH$ and the $qg\to qH$ processes, with respect to their collinear behavior. 
The latter is used to identify an interval of transverse momenta where the collinear approximation of the squared matrix element is accurate and where it is thus safe to apply the resummation procedure; 
we introduce a scale $w$ that represents the upper bound of this interval.
We discuss both scalar and pseudoscalar Higgs boson production
and determine, as a function of the Higgs and the quark masses,
in a model independent way, 
the scale $w$, which constitutes our main result.
In section~\ref{sec:smmass} we perform a phenomenological analysis in the SM: the numerical results of the previous section are applied in the analytic resummation context, 
with the code \hres,
and in the NLO+PS Monte Carlo framework,
with the code {\tt gg\_H\_quark-mass-effects} present in the \powhegbox; 
the corresponding Higgs $\pth$ distributions are eventually compared.
Finally, in section~\ref{sec:bottdom} we discuss the implications of the determination of the scale $w$
in the MSSM and in the 2HDM, with the possible production 
of new heavy states with masses of several hundred GeV and with a possible strong coupling of the Higgs to the bottom quark, enhanced with respect to the SM case.
For this study we use the generators {\tt gg\_H\_MSSM} and {\tt gg\_H\_2HDM}, also present in the {\tt POWHEG-BOX}.


%
 \section{Remarks on the computation of the Higgs \texorpdfstring{$\pth$}{ptH}  distribution}
\label{sec:generalresumm}
\subsection{Analytic resummation and the collinear limit}
\label{subsec:analyticresumm}
The Higgs boson acquires a transverse momentum $\pth$ because of its recoil against QCD radiation. 
In fixed-order perturbation theory the emission of initial state massless partons yields,
in the collinear limit, a logarithmic divergence of the Higgs transverse momentum distribution, signaling a breakdown of the perturbative approach, with an effective expansion parameter $\alpha_s(\pth) \log(\pth/m_H) \sim 1$ 
in the phase space region of vanishing $\pth$.
The analytic resummation to all orders of the terms  
$\left(\alpha_s(\pth) \log(\pth/m_H) \right)^n$
is performed by exploiting the universal properties of QCD radiation in the collinear limit and restores an acceptable physical behavior (the Sudakov suppression)
of the Higgs transverse momentum distribution in the limit $\pth\to 0$
\cite{Dokshitzer:1978hw,Parisi:1979se,Curci:1979bg,Collins:1981uk,Collins:1981va,Kodaira:1981nh,Kodaira:1982az,Altarelli:1984pt,Collins:1984kg,Catani:2000vq}.

In the collinear limit $\pth\to 0$ 
the amplitude for the real emission processes $gg\to gH$ and $qg\to qH$ diverge 
and can be written, via a Laurent expansion, as 
$\mathampl_{\text{exact}} = \mathampl_{\divtext}/\pth + \mathampl_{\regtext}$.
In this limit, the second term can be neglected with respect to the first one
and it is possible to recognize that $\mathampl_{\divtext}$ is proportional to the Born amplitude times the appropriate radiation term.
This factorized structure of the amplitude, neglecting the contribution coming from  $\mathampl_{\regtext}$ which is assumed to be small,
can be extended to all orders and it forms the  basis of the resummation procedure.
Indeed, we can iterate this factorization in the case of the amplitude for the emission of $n$ additional partons. 
In impact parameter space, this procedure leads to a factorized expression with $n$ divergent emission factors times a term proportional to the Born amplitude.
The expression for the approximated amplitude describing the emission of up to $n$ partons can be cast in the form of an exponential series, 
which can thus be summed to all orders.
The relative contribution of $\mathampl_{\regtext}$ to the full amplitude can be used to assess the accuracy of the collinear approximation and of the factorization hypothesis.

The resummed partonic cross section has a factorized structure given 
by the product of a universal exponential factor,
which accounts for the resummation to all orders of the logarithmically divergent terms,
multiplied by a process dependent function, 
which describes the details of the hard scattering process.
This factorization requires the introduction of a scale $\mu_{\text{res}}$, called resummation scale \cite{Bozzi:2003jy}.
The latter defines the region where the resummation is applied
and it is usually set to a value between 0 and the hard-scattering scale. 
A customary choice in the literature, for inclusive Higgs production, is to set the central value
 $\bar\mu_{\text{res}}=\mh/2$ \cite{Bozzi:2003jy}.
The precise choice of this value is one of the main topics of this paper and
will be further discussed in the next sections.
Analogously to what happens with the renormalization and factorization scales, 
the physical observables should not depend upon $\mu_{\text{res}}$,
but the truncation at a fixed order of the logarithmic expansion
leaves a residual dependence on it,
which can be used to estimate the uncertainty due to the missing higher-order logarithmic terms;
a variation of the scale $\mu_{\text{res}}$ in the interval $[\bar\mu_{\text{res}}/2,2\bar\mu_{\text{res}}]$ is customarily adopted.

The matching procedure requires to fix the integral of the Higgs transverse momentum distribution to a constant, 
which is conventionally set to the value of the fixed order total cross section \cite{Bozzi:2003jy}. 
This constraint holds exactly for any choice of  $\mu_{\text{res}}$, 
so that any variation of the resummation scale
modifies the shape of the distribution but not its integral 
and yields thus a correlation between low- and intermediate-$\pth$ regions.

\subsection{Numerical resummation in the NLO+PS framework}
\begin{figure}
  \centering
  \includegraphics[width=\textwidth]{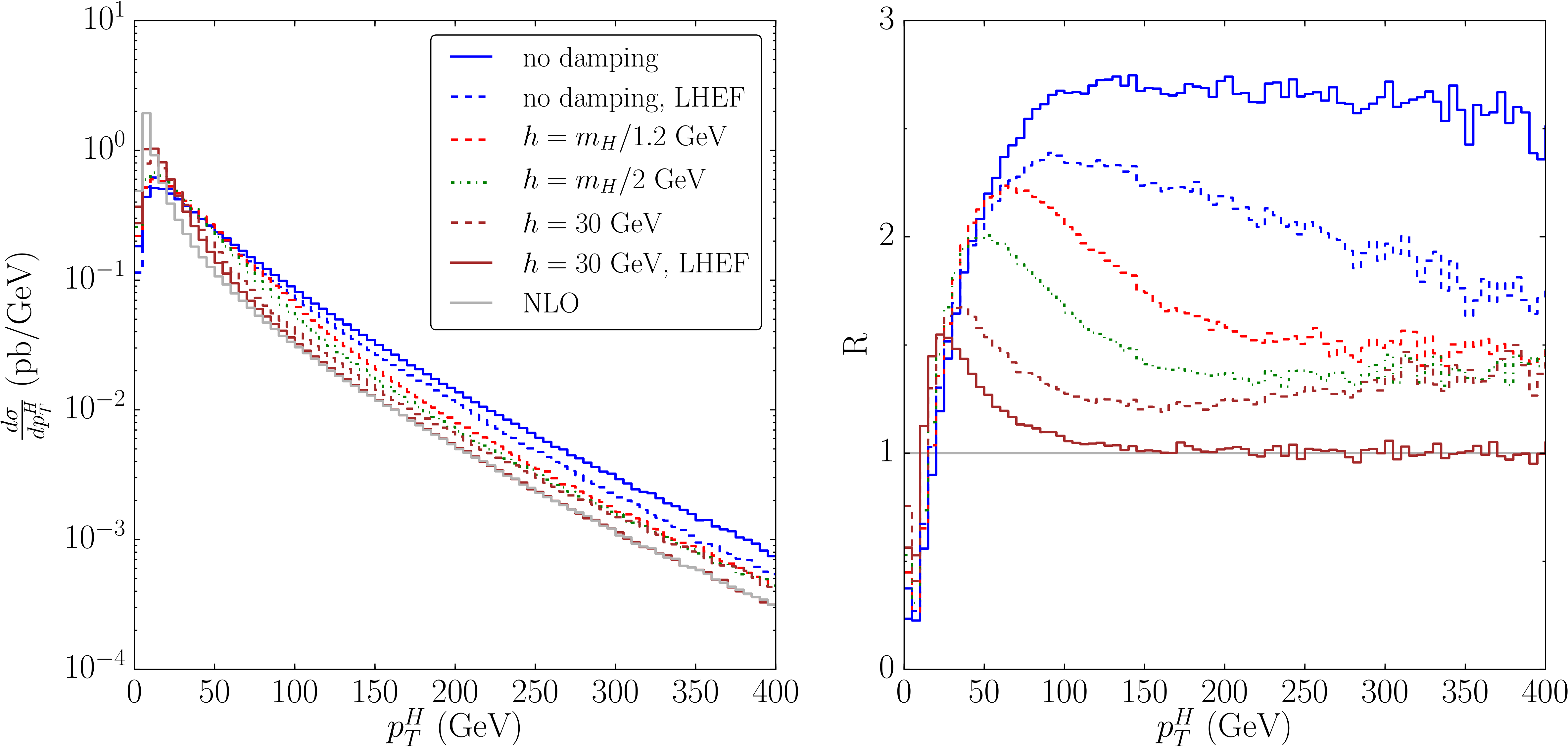}
  \caption{Left: effect of the damping factor $D_h$ for different values of the scale $h$ on the transverse momentum distribution of a SM Higgs of mass equal to $125$ GeV.
    The red dashed line is obtained with $h=m_H/1.2$ GeV, the green dot-dashed one with $h=m_H/2$ GeV and the indigo dashed one with $h=30$ GeV. The blue continuous line corresponds to no damping.
    For the no damping case and for $h=30$ GeV we also show the results at the level of Les Houches Event File (LHEF).
    For reference we show the NLO curve in gray. Right: ratio of the \powheg~prediction for the transverse momentum over the NLO result. The color coding is the same as in left figure.}
  \label{fig:2.1:hfactscan}
\end{figure}
\label{subsec:numresumm}
Another approach to the resummation of terms enhanced by the factor $\log(\pth/m_H)$ is the one obtained in the context of PS Monte Carlo, where the multiple emission
of partons is numerically simulated via the PS algorithm.
The matching between the fixed order NLO-QCD results and the PS has been
discussed in refs.~\cite{Frixione:2002ik,Frixione:2007vw,Alioli:2010xd} and it is implemented in several tools regularly used in the experimental analyses.

In a sufficiently general way we can write the matching formula as
\begin{align}
  d\sigma = \bar{B}^s(\Phi_B) d\Phi_B \left\{ \Delta^s_{t_0} + \Delta^s_t \frac{R^s(\Phi)}{B(\Phi_B)} d\Phi_r  \right\} + R^f d\Phi + R_{\text{reg}} d\Phi.
  \label{eq:sec2:matching}
\end{align}
The phase space is factorized into the product of the Born and the real emission components, $d\Phi = d\Phi_B d\Phi_r$.
The Born squared matrix element is denoted by $B$ while $\bar{B}$ is the NLO normalization factor. The latter is defined as
\begin{align}
  \bar{B}^s(\Phi_B) = B(\Phi_B) + \hat{V}_{\text{fin}} (\Phi_B) + \int \hat{R}^s (\Phi_B,\Phi_r) d\Phi_r\, .
\end{align}
In this formula $\hat{V}_{\text{fin}}$ represents the UV- and IR-regularized virtual contribution.
We use the hat to indicate that an amplitude has been IR-regularized.
The partonic subprocesses with the emission of an additional real parton
can be split into two groups: those that are divergent in the limit of collinear emission, called $ R_{\text{div}}$, and the ones that are instead regular, $R_{\text{reg}}$.
We can further subdivide the squared matrix elements of the divergent subprocesses in two parts:
\begin{align}
  R_{\text{div}} = R^s + R^f.
\label{eq:splitreal}
\end{align}
The term $R^s$ contains the collinear singularity of $R_{\text{div}}$, while $R^f$ is a finite remainder.
Finally, we use the symbol $\Delta_t^s$ for the Sudakov form factor, with $t$ as the shower
ordering variable:
\begin{align}
  \Delta^s_{t} = e^{-\int \frac{dt'}{t'} \frac{R^s}{B} d\Phi_r \theta(t'-t)}\, .
\label{eq:sudakov}
\end{align}
The splitting of $R_{\text{div}}$ in eq.~(\ref{eq:splitreal}) is defined up to a finite part which can be reabsorbed in $R^s$.
In the literature two different choices have been adopted:
in {\tt POWHEG} $R^s = R_{\text{div}}$, while
in {\tt MC@NLO} $R^s \propto \alpha_s P_{ij} B$ is proportional
to the product of the Born matrix elements times the relevant Altarelli-Parisi splitting functions.

It is interesting to observe that different definitions for $R^s$ 
generate higher-order effects in the matched differential cross section.
The possibility of defining the finite part $R^f$ in an arbitrary way can be exploited to parameterize the uncertainties related to the matching procedure.

\subsubsection{The role of the damping factor \texorpdfstring{$D_h$}{D_h} in the \powhegbox~framework}

In the \powhegbox~framework, the separation between $R^s$ and $R^f$ can be achieved in a dynamical way using the damping factor $D_h$, defined as
\begin{align}
  D_h = \frac{h^2}{h^2+(p_{\bot}^H)^2}\,.
  \label{eq:resframework:dh}
\end{align}
The divergent and the regular part of $R_{\text{div}} = R^s + R^f$ are then 
defined as:
\begin{align}
  R^s = D_h~R_{\text{div}} ~,\phantom{aaaaaaaaaaa} R^f = \left(1-D_h\right)~R_{\text{div}}\,.
\end{align}
The role of the scale $h$ is to separate
the low and the high transverse-momentum regions and it therefore specifies the range of momenta for which the Sudakov form factor is possibly different from $1$.
In the limit
$\pth \ll h$  we obtain
$R^s \rightarrow R_{\text{div}}$ and $R^f \rightarrow 0$.
In this limit the Higgs $\pth$ distribution is suppressed by the Sudakov form factor.
On the other hand, when $p_{\bot}^H \gg h$ we have $R^s \rightarrow 0$ and $R_f \rightarrow R_{\text{div}}$ and the Sudakov form factor tends to $1$.
In this latter regime
the emission of a real parton is described at fixed order by the matrix elements $R^f = R_{\text{div}}$.

The differential distribution generated according to eq.~(\ref{eq:sec2:matching}) contains higher order terms,
beyond the claimed accuracy of the calculation,
due to the product of $\bar{B} \times R^s$.
Indeed in the large $\pth$ region we have
\begin{align}
  d\sigma &= \bar{B}(\Phi_B) d\Phi_B \left\{ \Delta_{t_0} + \Delta_t \frac{R^s(\Phi)}{B(\Phi_B)} d\Phi_r \right\} + R^f d\Phi + R_{\text{reg}} d\Phi \nonumber \\
&\approx \bar{B}(\Phi_B) \frac{R^s(\Phi)}{B(\Phi_B)} d\Phi + R^f d\Phi + R_{\text{reg}} d\Phi \nonumber \\
&\equiv K(\Phi_B) R^s(\Phi) d\Phi + R^f d\Phi + R_{\text{reg}} d\Phi, \nonumber \\
K(\Phi_B) &\equiv \frac{\bar{B}(\Phi_B)}{B(\Phi_B)} = 1 + \mathcal{O}(\alpha_s)\,.
\end{align}
Originally the factor $D_h$ was introduced to damp the $R^s$ contribution at large $\pth$ and to recover
the exact fixed order result in this kinematic region, at the level of the first emission handled by \powheg.

By varying the scale $h$, it is possible to check how well the fixed order distribution is recovered for large values of $\pth$, as can be seen from figure~\ref{fig:2.1:hfactscan}.

We observe that, while at the level of the first emission generated by POWHEG (obtained at the level of Les Houches Event File (LHEF))
the NLO result is fully recovered, the showering of the events causes the high-$\pth$ tail of the distribution to rise over the NLO prediction.

The total NLO cross section is always preserved for any value of $h$, as can be checked by integrating eq.~(\ref{eq:sec2:matching}) over the whole phase space.
This property implies in turn that the low- and high-$p_{\bot}^H$ regions of the differential cross section are correlated.
Any increase of the distribution at low-$p_{\bot}^H$ translates in a decrease of the high-$p_{\bot}^H$ tail and vice versa.

The role effectively played by the scale $h$ has some similarities with the one described in section \ref{subsec:analyticresumm} for the resummation scale $\mu_{\text{res}}$:
indeed, for $\pth<h$ or for $\pth<\mu_{\text{res}}$
the Sudakov suppression yields a regular behavior of the Higgs transverse momentum distribution, whereas for $\pth$ larger than these scales the fixed-order description is recovered, at the level of description given
by \powheg.
It should however be remarked that $\mu_{\text{res}}$ and $h$
have a completely different origin.
The scale $\mu_{\text{res}}$ is introduced
as the scale at which the resummation is defined and the factorization of the partonic cross section implemented.
It necessarily appears in the arguments of the logarithmic terms that are resummed.
The damping factor $D_h$ is instead a convenient $\pth$-dependent parameterization of the ambiguity in the definition of $R^s$.
In a different perspective, the scale $h$ controls the range of $\pth$ over which
the first term in eq.(~\ref{eq:sec2:matching}) is active in the generation of the first real emission.
Since this term contains the normalization factor $\bar B$,
the scale $h$ in turn controls also how the total NLO cross section is spread over the $\pth$ distribution.

\subsection{The value of the {\tt SCALUP} variable}
\label{sec:scalup}
The emission of the radiation in the \powheg~approach is described by Eq.~\ref{eq:sec2:matching}.
Neglecting the contribution coming from the term $R_{reg}$ 
(negligible in the case of the Higgs production in gluon fusion), 
we have two different categories of events. 
One corresponds to the terms in curly brackets ($\bar{B}$-events), 
while the second one is described by the term $R^f$ (remnant events). 
The latter is present only if the damping factor $D_h$ is used.

To avoid double counting of the emissions, 
in the \powheg~approach the PS is required to emit radiation at transverse-momentum scale lower than the one of the parton emitted by the \powhegbox. 
More in detail, in the case of $\bar{B}$ events the PS should start to consider the possibility of an emission exactly at the scale
at which the \powheg~parton was radiated. 
In the default \powhegbox~implementation the same choice is applied,
for a reason of uniformity, also to the remnant events.
This information is therefore computed on an event-by-event basis and the passed to the PS using the {\tt SCALUP} field in the LHE event record.

It might happen that the value of {\tt SCALUP} is large and that the description by the PS of real radiation at large transverse momenta is not accurate,
since this approach is based on the soft/collinear approximation.
It is then natural to consider as an option the possibility of setting an upper bound to the value of the {\tt SCALUP} variable,
close or equal to the one adopted for the $h$ parameter\footnote{more precisely, we take $\min(\pth,h)$}.
This choice is applied only to the events generated by the $R^f$ part of the real matrix elements, in order to respect the \powheg~accuracy given by the first term in Eq.~\ref{eq:sec2:matching}
and account for a higher-order effect. 
Since these events are relevant only for the description of the high-$\pth$ region (in turn defined by the scale choice $h$), 
we do not expect a modification of the shape of the distribution in the low-$\pth$ region. 
Indeed, in this way the action of the PS is restricted to the lower part of the $\pth$ spectrum,
whereas the large-$\pth$ tail is described purely by the LO matrix elements.
In Section~\ref{sec:smmass}  we will compare the description of the Higgs high-$\pth$ tail with the default and with the modified {\tt SCALUP} values, in the case of the SM.


%
 \section{Collinear approximation of partonic squared matrix elements}
\label{sec:hggcoll}

In the previous section we have recalled 
that the resummation to all orders of the terms enhanced by a 
$\log(\pth/\mh)$ factor is possible
thanks to the factorization of the squared matrix element in the collinear limit $\pth\to 0$.
Based upon those considerations, 
we now explain our procedure to determine the accuracy of the collinear approximation of the full squared matrix element
with respect to $\pth$,
focusing for simplicity on the channel $gg \to gH$.
We then derive numerically the value $w$ of the upper limit of the $\pth$ range where the collinear approximation is accurate,
for the scalar and the pseudoscalar final states, 
considering both the $gg \to gH$ and the $qg \to qH$ channels.

\subsection{Helicity amplitudes and kinematic variables}
We consider the helicity amplitudes\footnote{
${\lambda_1 = \pm 1}$, $\lambda_2 = \pm 1$ are the helicities of the two incoming gluons and $\lambda_3 = \pm 1$ is the helicity of the outgoing one.} 
$\mathampl^{\lambda_1,\lambda_2,\lambda_3}(s,\pth,m_H^2)$ for the process $gg \to gH$, 
whose complete expressions can be found for example in ref.~\cite{Baur:1989cm}.
We reorganize them, via a Laurent expansion, as follows:
\begin{align}
  \mathampl^{\lambda_1,\lambda_2,\lambda_3}(s,\pth,m_H^2) = \mathampl_{\divtext}^{\lambda_1,\lambda_2,\lambda_3}(s,m_H^2)/\pth + \mathampl^{\lambda_1,\lambda_2,\lambda_3}_{\regtext}(s,\pth,m_H^2)
\end{align}
and we use this decomposition to compute the unpolarized squared matrix element exactly, $|\mathampl|^2$, or its collinearly divergent part
$|\mathampl_\divtext/\pth|^2$.

We define the ratio $C$:
\begin{align}
  C(s,\pth,m_H^2) = \frac{|\mathampl(s,\pth,m_H^2)|^2}{|\mathampl_{\divtext}(s,m_H^2)/\pth|^2}\, ,
\end{align}
which quantifies how the unpolarized exact squared matrix element differs from its collinear approximation as a function of $\pth$.
We observe that by construction we have $\lim_{\pth \rightarrow 0} C(s,\pth,m_H^2) = 1$.
In our study we also consider the behavior of the interference term between the top and the bottom quark. 
For this specific case we redefine the parameter $C$ as
\begin{align}
  C_{\text{int}}(s,\pth,m_H^2) = \frac{2 \text{Re} \left( \mathampl_t(s,\pth,m_H^2) \mathampl^{*}_b(s,\pth,m_H^2) \right) }{2 \text{Re} \left( \mathampl_{\divtext,t}(s,\pth,m_H^2) \mathampl^{*}_{\divtext,b}(s,\pth,m_H^2) \right)/\left(\pth\right)^2}\,.
\end{align}

We introduce the following practical criterion:
the regular part of the amplitude becomes non-negligible with respect to its collinear counterpart for a value $w$ of $\pth$ such that 
\begin{align}
\left|C(s,w,m_H^2)-1 \right| > \bar C\,.
\label{eq:cbar}
\end{align} 
To fix the setup of our study we choose $\bar{C} = 0.1$. This value is arbitrary, but its order of magnitude can be justified in the framework of a QCD calculation, since the size of the terms without a collinear logarithmic enhancement is $\alpha_s/\pi$ times a coefficient of order 1. We do not assign any special meaning to the scale that will be found with our analysis but we rather consider it as a starting point to compute an uncertainty band. 
At the end of the section we analyze the dependence on the specific value of $\bar{C}$.

The amplitude of the process $gg\to H g$ is a function of two independent kinematic variables, e.g. $s$ and $\pth$.
The production of a final state with a definite $\pth$ requires a minimum value for $s$:
\begin{align}
  s_{\text{min}} = m_H^2 + 2 (\pth)^2 + 2 \pth \sqrt{ (\pth)^2 + m_H^2 }\,.
  \label{eq:helampl:smin}
\end{align}
We study the behavior of the amplitude as a function of $\pth$ for $s = s_{\text{min}} + s_{\text{soft}}$,
where $s_{\text{soft}}$ is necessary to avoid the soft divergence and focus only on the collinear behavior.
The choice of a value of $s$ close to $s_{\text{min}}$ is phenomenologically motivated by the strong PDF suppression in the hadronic cross section
for increasing partonic $s$.

An analogous procedure is used to determine the scale $w$ for the $qg \to qH$ subprocess,
with the analytic expressions of ref.~\cite{Bonciani:2007ex} 
and in the case of pseudoscalar production using the formulae in ref.~\cite{Degrassi:2011vq}.

\subsection{Partonic analysis}
We assume that the full amplitude is the sum of a top and a bottom contribution, neglecting the light quark generations.
Furthermore, we do not consider the possibility of additional colored particles running in the loop, since the current LHC results hint to the fact that, if these states exist, their mass is probably much larger than the top mass.
Therefore they would not affect the shape of the Higgs transverse momentum distribution for values of $\pth$ that are phenomenologically interesting.

Under the assumption that the coupling of the gluons to the quarks is the one dictated by QCD and 
that all the details about the coupling of the Higgs to the quarks can be factorized from the rest of the amplitude, 
we can consider the value of the scales $w_t$, $w_b$ and $w_i$, that we respectively find  
in the case of squared matrix elements with only top quark diagrams, only bottom diagrams or for the top-bottom interference,
as model independent.
As a consequence, 
the determination of the scales depends only on the quark and the Higgs masses.

While the scales computed with only one quark might have a physical interpretation in the BSM scenarios where  that quark yields the dominant contribution to the cross section,
the scale of the interference terms, while unphysical, is a necessary tool to treat accurately the full theory in scenarios where both top and bottom quarks are equally important. Since the full squared matrix element, including top and bottom quarks, factorizes in the collinear limit, the same pattern should be followed not only by the terms with the squared amplitude of one single quark, but also by the interference terms, making our treatment viable.

In sections \ref{sec:smmass} and \ref{sec:bottdom}
we will discuss how these results can be exploited in a model specific framework.

\subsubsection{Scalar Higgs}

\begin{figure}
  \centering
  \includegraphics[width=0.49\textwidth]{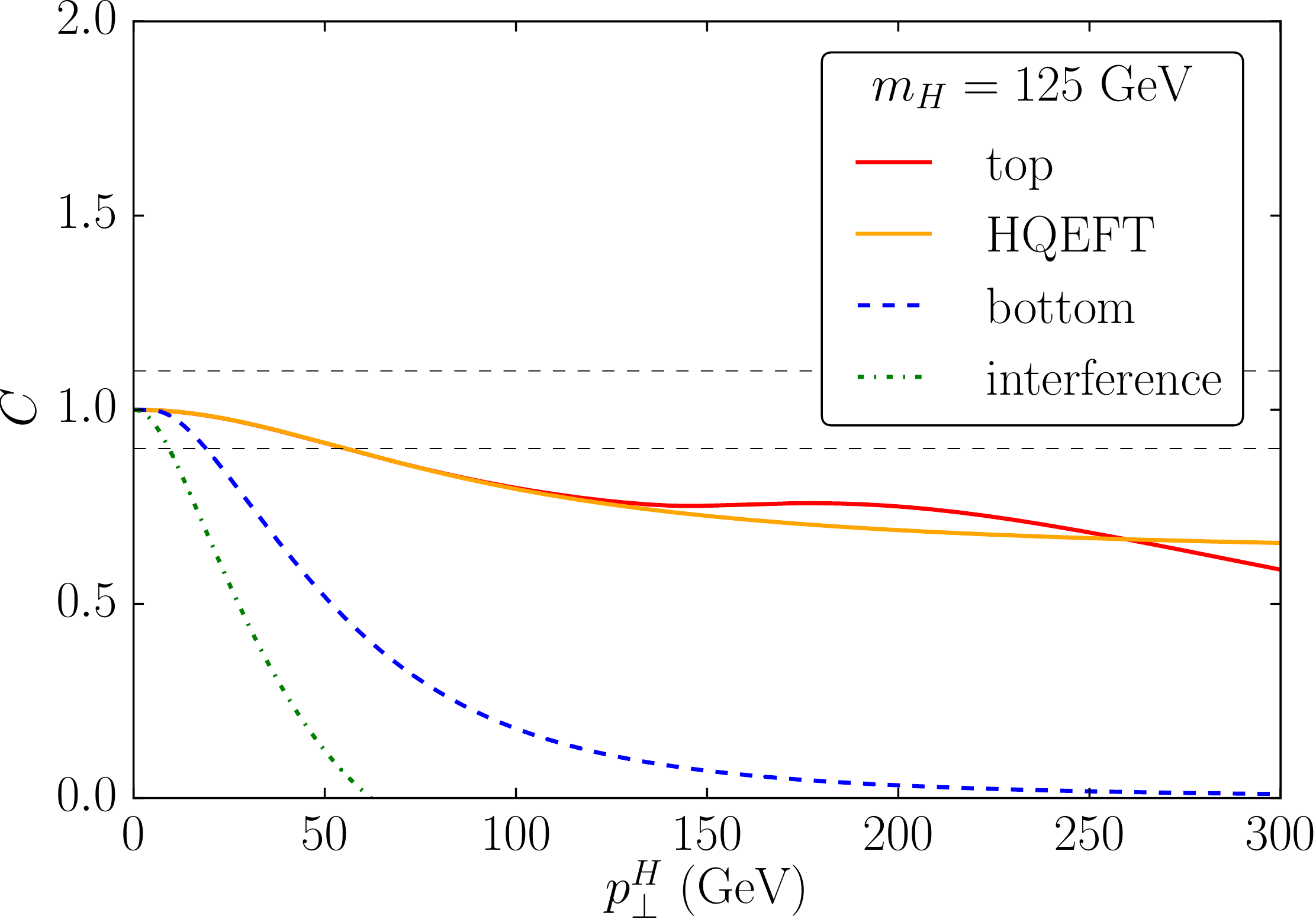} \includegraphics[width=0.49\textwidth]{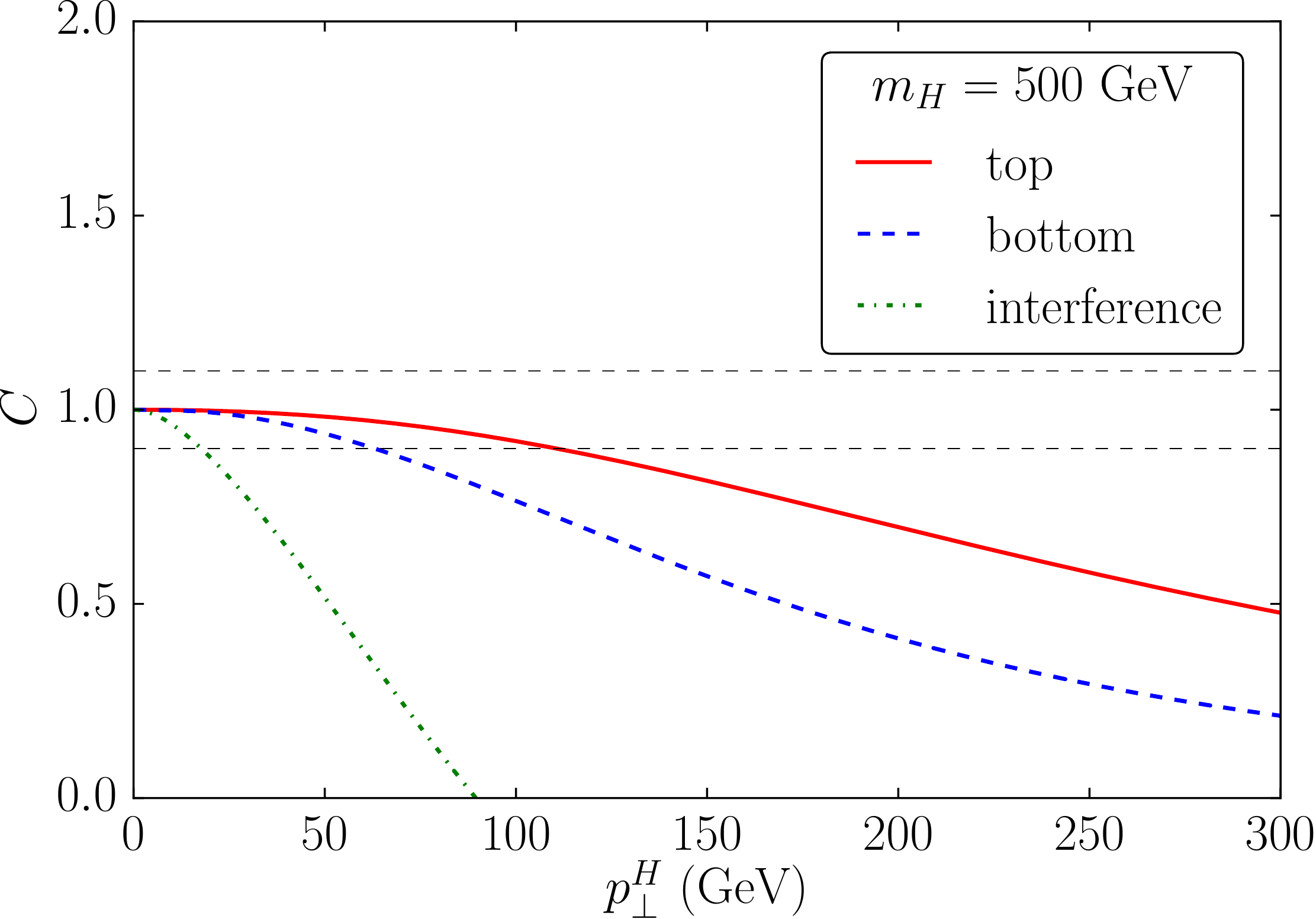}
  \caption{Relative effect of the regular part of the amplitude compared to the collinear approximation, for a light Higgs (left, $m_H = 125$ GeV) and for a heavy Higgs (right, $m_H = 500$ GeV), in the $gg$ channel.
    In red we show the results for the squared top quark amplitude, in blue the ones for squared bottom amplitude and in green the ones for the interference. For comparison, in the case of $m_H = 125$ GeV,
      we also plot     the curve for the HQEFT in orange.
}
  \label{fig:amplcoll:Cpth1}
\end{figure}

To exemplify the outcomes of our procedure, we show the results for the variable $C(s,\pth,m_H^2)$ for a Higgs boson with $m_H = 125$ GeV and $m_H = 500$ GeV in figure~\ref{fig:amplcoll:Cpth1},
in the case of the $gg \to gH$ subprocess.
We plot in red and blue the behavior of the squared matrix elements 
computed including only the top or only the bottom diagrams,
in green we show the behavior of the 
interference of the top and bottom amplitudes.
In the same figure, for $m_H = 125$ GeV, we plot in orange the results obtained by applying the same procedure to the HQEFT matrix elements.

We first discuss the impact of the regular terms in the case of a light Higgs.
We compare the results obtained with the exact matrix elements including only the top quark with the ones in the HQEFT;
we observe that in both models a deviation by more than 10\% from the collinear approximation occurs for $\pth>55$ GeV. 
Since it is present in both cases, this effect should thus not be interpreted as a top mass effect; the latter becomes visible for $\pth> 150$ GeV.
From the analysis of the helicity amplitudes, we observe that this deviation from the collinear approximation stems from $\mathampl^{-+-}$.
For the bottom quark, the deviation from the collinear approximation starts from $\pth > 19$ GeV.
In the case of the interference terms, we observe that the determination 
of the scale $w_i$ is dominated by the behavior of the bottom amplitude; the corresponding value, $w_i=9$ GeV, 
is smaller than the ones obtained in the other two cases.


In the case of a heavy Higgs, with $\mh > m_t, m_b$, the scale of the process is set by the mass of the boson (e.g. $\mh=500$ GeV)
and the HQEFT approximation of the amplitude is not valid.
We observe that the amplitude that includes only the top-quark diagrams deviates from its collinear approximation\footnote{
We remark that the collinear regime is not defined by a given value 
$\overline \pth$ of the variable $\pth$, defined as the value at which the 
deviation from the collinear behavior is equal to $\bar{C}$, but it is better characterized in terms of the ratio $r=\overline{\pth}/\mh$; 
in the case under discussion (only top diagrams) we find $r\simeq 1/4$ whereas for $\mh=125$ GeV we have $r\simeq 1/2$.}
for $\pth > 111$ GeV.
Instead,  the squared matrix element that includes only the bottom-quark diagrams deviates from its collinear approximation
for $\pth > 63$ GeV. 
Finally, for the interference terms we find the bound $\pth > 18$ GeV.

\begin{table}[!h]
\centering
\begin{center}
\begin{tabular}{@{}ccccccccc|ccc}
\toprule
& \multicolumn{11}{c}{\bf Scalar, collinear deviation scale $w$ (GeV)}\\
\midrule
$\mh$ (GeV) & & $w^{gg}_t$ & $w^{gg}_b$ & $w^{gg}_i$ & & $w^{qg}_t$ & $w^{qg}_b$ & $w^{qg}_i$ & $w^{gg+qg}_t$ & $w^{gg+qg}_b$ & $w^{gg+qg}_i$ \\
\cmidrule{1-1} \cmidrule{3-5} \cmidrule{7-9} \cmidrule{10-12}
$125$ & & $55$ & $19$ & $9$ & & $24$ & $7$ & $5$ & $48$ & $18$ & $9$ \\
$200$ & & $85$ & $29$ & $16$ & & $21$ & $5$ & $5$ & $71$ & $27$ & $14$ \\
$300$ & & $132$ & $41$ & $25$ & & $17$ & $4$ & $4$ & $111$ & $38$ & $23$ \\
$350$ & & $102$ & $47$ & $28$ & & $15$ & $4$ & $4$ & $87$ & $43$ & $26$ \\
$400$ & & $94$ & $52$ & $26$ & & $14$ & $4$ & $3$ & $81$ & $49$ & $23$ \\
$500$ & & $111$ & $63$ & $18$ & & $13$ & $3$ & $2$ & $96$ & $58$ & $17$ \\
$600$ & & $133$ & $73$ & $6$ & & $13$ & $3$ & $0$ & $113$ & $68$ & $6$ \\
$700$ & & $157$ & $83$ & $25$ & & $9$ & $2$ & $2$ & $137$ & $78$ & $24$ \\
$800$ & & $181$ & $93$ & $46$ & & $8$ & $2$ & $36$ & $158$ & $87$ & $46$ \\
\bottomrule
\end{tabular}
\end{center}
\caption{Value of the scales $w_{t,b,i}$ for a scalar Higgs. The scales are reported both as determined separately in the two partonic subprocess (left) and after their combination according to eq.~\ref{eq:weightedcomb} (right).}
\label{tab:hthbscalar}
\end{table}

\begin{figure}
  \centering
  \includegraphics[width=0.90\textwidth]{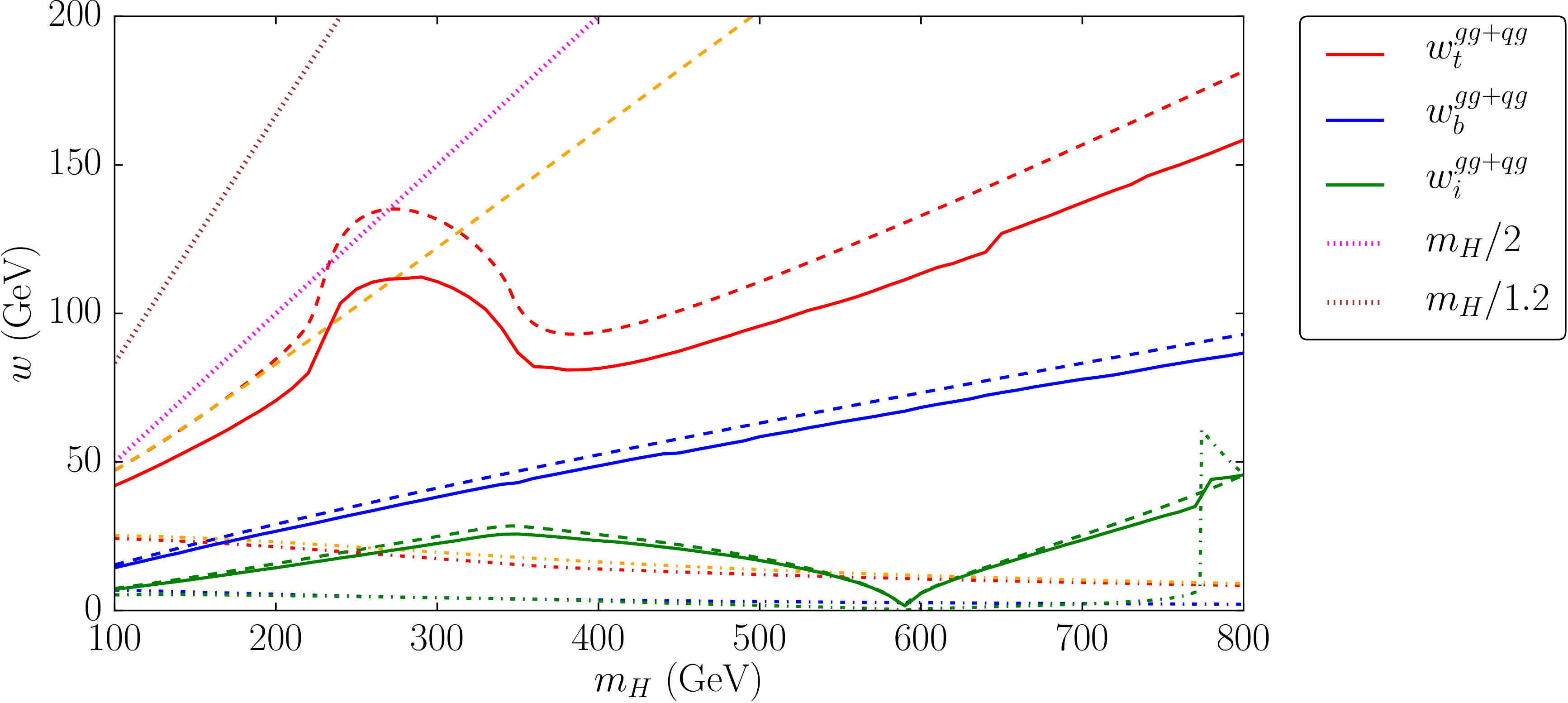} 
  \caption{Combination of the scales $w^{gg}$ and $w^{qg}$ according to eq.~(\ref{eq:weightedcomb}). In red we show the result for the top quark, in blue the one for the bottom and in green the one for the interference term.
    The dashed style represents the scales obtained in the $gg$ channel, the dot-dashed style the ones in the $qg$ channel. Continuous lines are used for the merged scales. 
We also show as a dotted line the scale choices $m_H/2$ 
\protect\cite{Bozzi:2003jy}
and $m_H/1.2$ 
\protect\cite{Dittmaier:2012vm}.
In orange we show the results for the HQEFT.
  \label{fig:hthbhiscalar}}
\end{figure}

In the left section of table~\ref{tab:hthbscalar} and in figure~\ref{fig:hthbhiscalar}
we present the values of the scales $w$, 
derived from the study of scalar Higgs production
for different choices of $\mh \in [125,800]$ GeV, 
separately in the case of
squared matrix elements computed including only the top, only the bottom diagrams or for the interference of the top and bottom amplitudes;
the results are presented separately for the two partonic subprocesses,
$gg\to gH$ and $qg\to qH$. A finer scan in $\mh$ is available in table~\ref{tab:wtwbwiappendix1} in the appendix.

We observe that in the $gg\to gH$ channel both scales $w_{t,b}$ increase with the Higgs mass,
with the exception of the region of real top-pair production threshold, where the effect on $w_t$ of additional terms that induce a deviation from the collinear approximation is visible. In the bottom-quark case such phenomenon does not show up, because for realistic values of $\mh$ the process scale is always well above the bottom-pair production threshold.
The interference scale $w_i$ has a peculiar behavior: in fact, it shows a growth with $\mh$ until the top-pair production threshold and then it decreases for larger $\mh$, until it vanishes for $\mh=589$ GeV,
with our $\mt$ and $\mb$ choices;
for even larger $\mh$ values it grows again.
In order to explain why $w_i$ vanishes, we should recall that 
the interference terms, as a function of $\mh$ and for fixed $m_t$ and $m_b$, are not positive definite and may change sign for a specific value of $\mh$;
in particular, when the underlying LO (i.e. of the process $gg\to H$) interference terms vanish, also the collinear approximation does. In this point the interference terms of the processes $gg\to gH$ and $qg\to qH$ are thus collinear finite,  the function $C(s,\pth,\mh)$ diverges for all $\pth$ and the scale $w_i$ is equal to zero, indicating that the $\pth$ distribution is regular and a LL resummation is not needed. It should be noted that, for this specific configuration,
the importance of the interference term is in any case small, since it vanishes at LO.

We observe that in the $qg\to qH$ channel the scales are lower than in the previous case
and that they decrease for increasing values of the Higgs mass.
The basic argument to explain this different behavior can be found analytically in the HQEFT: we expand the ratio $C(s,\pth,\mh^2)$
in powers of $\pth$ around $\pth=0$ and we find
\bea
C_{qg}^{HQEFT} &=& 1 - \mh^2 \,\frac{(\pth)^2 }{s_{soft}^2}\, \frac{\left(1+\frac{s_{soft}}{\mh^2}\right) \left(1+2\frac{s_{soft}}{\mh^2}+
4 \frac{s_{soft}^2}{\mh^4}\right)}
{\left(1+2\frac{s_{soft}}{\mh^2}+2\frac{s_{soft}^2}{\mh^4}\right)}
+{\cal O}((\pth)^3)\,, \\
C_{gg}^{HQEFT} &=& 1 - 2 \,\frac{(\pth)^2\,\, s_{soft}^2}{\mh^6} \frac{1+\frac{s_{soft}}{\mh^2}}
{\left(1+\frac{s_{soft}}{\mh^2}+\frac{s_{soft}^2}{\mh^4}\right)^2} 
+{\cal O}((\pth)^3)\,.
\eea
The different behavior with respect to $\mh$ of the scale $w$ is due,
in the $gg$ case, to the fact that the function $C$ receives corrections with negative powers of $\mh$, so that for heavy Higgs masses there is a larger interval of $\pth$ where the collinear limit provides a good approximation 
of the full result;
in the $qg$ case instead, there are corrections quadratic in $\mh$, such that the deviation of $C$ from 1, for large $\mh$, occurs at smaller $\pth$ values.
A numerical analysis with the full dependence on the top and bottom masses confirms the explanation derived above in the HQEFT.

The interference scale vanishes, as expected, 
for the same $\mh$ value
in the $gg\to gH$ and the $qg\to qH$ channel,
since they factorize to the same LO term.

The different values of the scales $w_{t,b}$ obtained in the two partonic channels $gg$ and $qg$
give rise to a practical problem, in case one wants to use at hadron level one single scale to control the effects of multiple parton emissions;
given that the $w$ value from the $gg$ channel is always larger than the one from the $qg$ channel, we can expect that the final value will lie in between;
we evaluate it with a weighted average, 
with the relative contributions of the two channels in each bin, 
further adjusted to account for the shape of the physical distribution. We define
\bea
w^{gg+qg}(\mh) &\equiv& 
\int_{w^{qg}}^{w^{gg}}d\pth
\left( w^{gg} \frac{\frac{d\sigma^{gg}}{d\pth}}{\frac{d\sigma^{gg+qg}}{d\pth}} + w^{qg} \frac{\frac{d\sigma^{qg}}{d\pth}}{\frac{d\sigma^{gg+qg}}{d\pth}}    \right)
\times
\frac{\frac{d\sigma^{gg+qg}}{d\pth}}{\sigma^{interval} }\,,
\label{eq:weightedcomb}
\eea
where
\bea
\sigma^{interval}&=&\int_{w^{qg}}^{w^{gg}}d\pth\, \frac{d\sigma^{gg+qg}}{d\pth}\,.
\eea
In figure~\ref{fig:hthbhiscalar}, and in table~\ref{tab:hthbscalar},
in the last three columns to the right, 
we show the results of this combination, 
which are our best determination for the scales to be used in the simulation 
of the hadronic differential cross section
\footnote{ The relative weight of the two partonic channels, as a function of the Higgs mass, is slowly varying, so that we can approximate  the result of equation \ref{eq:weightedcomb} with the simpler relations 
$w_t=0.2\, w^{qg}_t + 0.8\, w^{gg}_t$ and
$w_b=0.1\, w^{qg}_b + 0.9\, w^{gg}_b$. These relations approximate the exact combination at the 5\% level.}.
We have used the code {\tt SusHi}~\cite{Harlander:2012pb}, with $\sqrt{S}=13$ TeV, to compute the weights used in eq.~(\ref{eq:weightedcomb}).
Since this procedure requires the evaluation of the hadronic cross section, the combined scales are dependent on the $\sqrt{S}$ value used and on the
other hadronic parameters. In particular this is true for the choice of the renormalization and factorization scale, that we have assumed to be $\mu_r = \mu_f = m_H$.
However we have verified that the effect on the channel-combined value for the scales is only at the of few GeVs, well within the uncertainty band that we are considering.
A finer scan in the Higgs mass, is provided in tables~\ref{tab:wtwbwiappendix1} and \ref{tab:wtwbwiappendix2}, in Appendix \ref{app:hthbscan}.

\subsubsection{Pseudoscalar Higgs}
In table~\ref{tab:hthbpseudoscalar} 
we present a sample of the 
results,
analogous to the ones of the previous subsection,
for the case of pseudoscalar Higgs production.

\begin{table}[!h]
\centering
\begin{center}
\begin{tabular}{@{}ccccccccc|ccc}
\toprule
& \multicolumn{11}{c}{\bf Pseudoscalar, collinear deviation scale $w$ (GeV)}\\
\midrule
$m_A$ (GeV) & & $w^{gg}_t$ & $w^{gg}_b$ & $w^{gg}_i$ & & $w^{qg}_t$ & $w^{qg}_b$ & $w^{qg}_i$ & $w^{gg+qg}_t$ & $w^{gg+qg}_b$ & $w^{gg+qg}_i$ \\
\cmidrule{1-1} \cmidrule{3-5} \cmidrule{7-9} \cmidrule{10-12}
125 & & $60$ & $19$ & $11$ & & $24$ & $7$ & $6$ & $52$ & $18$ & $10$ \\
200 & & $126$ & $29$ & $18$ & & $22$ & $5$ & $5$ & $102$ & $27$ & $16$ \\
300 & & $122$ & $41$ & $28$ & & $18$ & $4$ & $4$ & $103$ & $38$ & $25$ \\
350 & & $82$ & $47$ & $25$ & & $15$ & $4$ & $4$ & $70$ & $43$ & $23$ \\
400 & & $99$ & $52$ & $15$ & & $14$ & $4$ & $2$ & $86$ & $49$ & $14$ \\
500 & & $127$ & $63$ & $15$ & & $12$ & $3$ & $2$ & $109$ & $58$ & $14$ \\
600 & & $155$ & $73$ & $36$ & & $11$ & $3$ & $51$ & $132$ & $68$ & $39$ \\
700 & & $184$ & $83$ & $69$ & & $10$ & $2$ & $18$ & $160$ & $77$ & $60$ \\
800 & & $212$ & $92$ & $277$ & & $9$ & $2$ & $10$ & $184$ & $86$ & $239$ \\
\bottomrule
\end{tabular}
\end{center}
\caption{Value of the scales $w_{t,b,i}$ for a pseudoscalar Higgs. The scales are reported both as determined separately in the two partonic subprocess (left) and after their combination according to eq.~\ref{eq:weightedcomb} (right).}
\label{tab:hthbpseudoscalar}
\end{table}

The general behavior of the two partonic channels is similar to the one observed for scalar production. One difference can be observed at the top-pair threshold, where a cusp appears in the $w_t$ prediction, 
reflecting the analogous feature of the total cross section.
The scale $w_i$ vanishes for a different value of the pseudoscalar mass, $m_A=445$ GeV, because of the different LO dependence on $m_A$, $\mt$ and $\mb$.
As for the scalar case, a more detailed scan as a function of $m_A$ is available in tables~\ref{tab:wtwbwiappendix1} and \ref{tab:wtwbwiappendix2}, in Appendix \ref{app:hthbscan}.

\subsection{Dependence on auxiliary parameters}
\label{sub:depaux}
\begin{figure}
  \centering
  \includegraphics[width=0.49\textwidth]{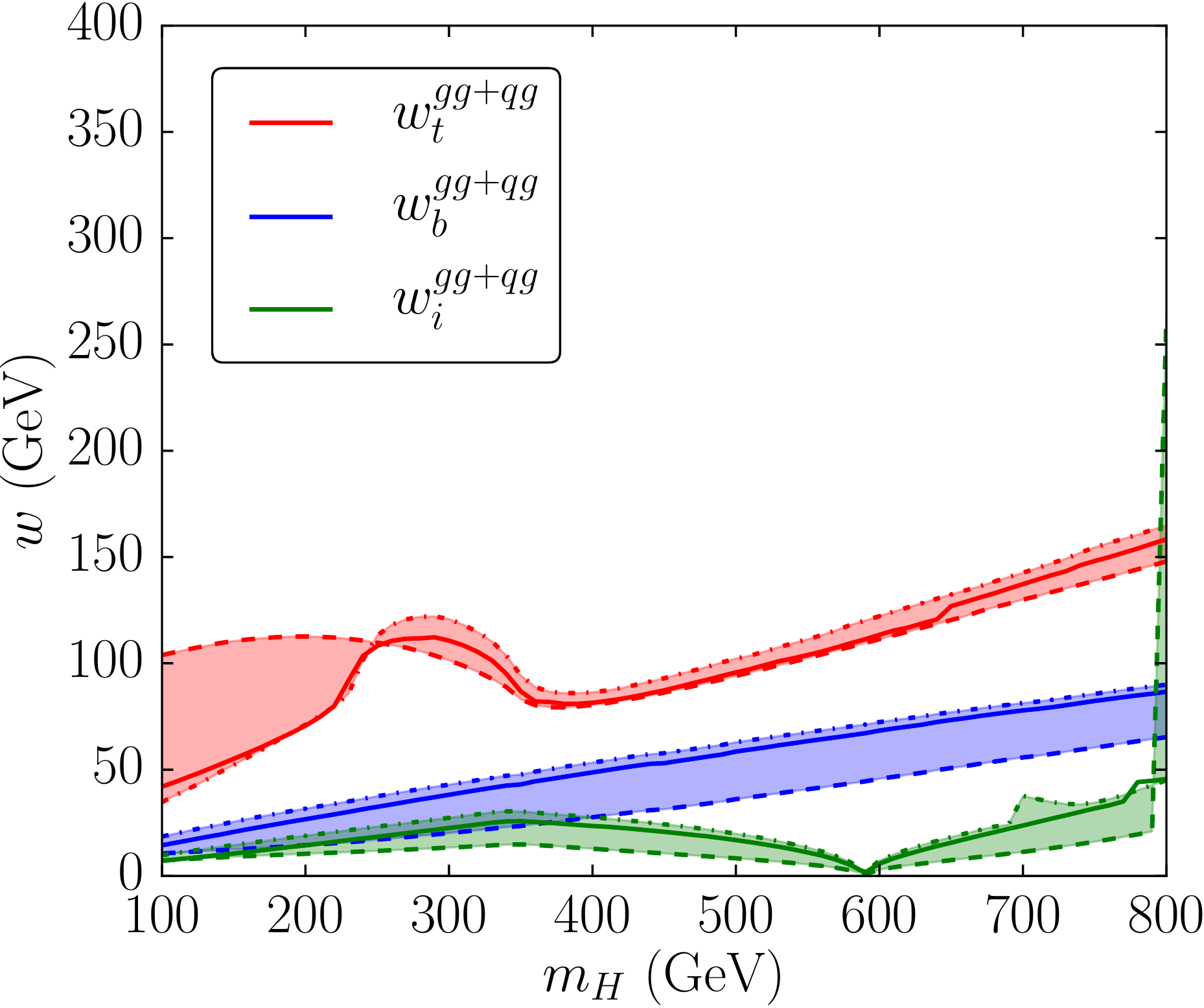}
  \includegraphics[width=0.49\textwidth]{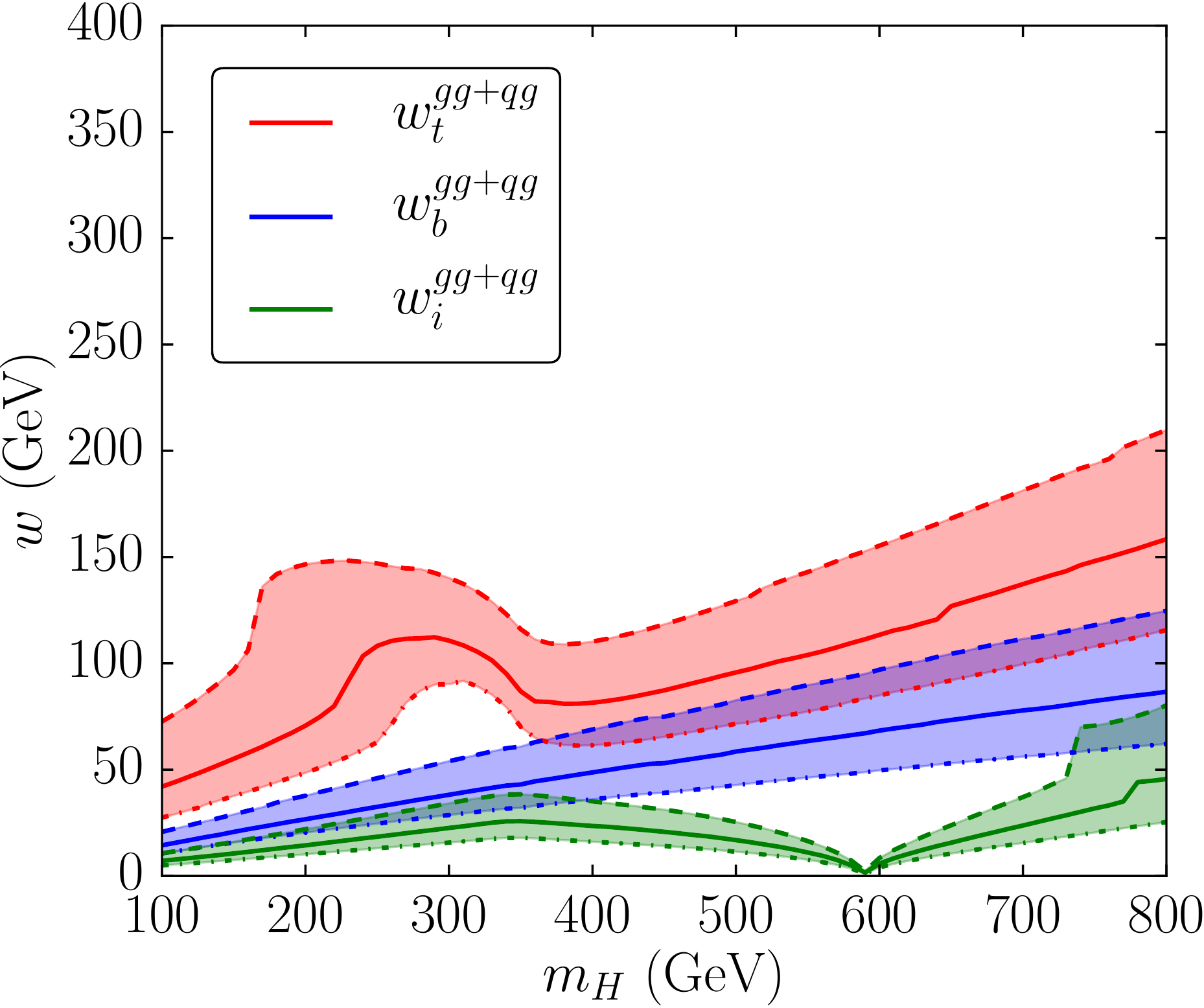} 
  \caption{Auxiliary parameter sensitivity for the merged $gg$-$qg$ scales. On the left, dependence of the scale determination on the choice of the value of the cut-off $s_{soft}$; on the right dependence of the scale determination on the choice of the value $\bar C \in [0.05,0.2]$. The dashed curves represent the values obtained by enlarging  the parameter whose dependence is under study while the dot-dashed curves are obtained by the rescaling of the parameter to a smaller value.
\label{fig:auxdepcombined}
}
\end{figure}

The value of the resummation scale has been determined with an analysis of the partonic squared matrix element, for fixed value of the partonic invariant $s$. 
For a given final state configuration and in particular for a given value of $\pth$, the hadronic distribution receives contributions from all the partonic cross sections with $s_{\text{min}}\le s \le S$, where $S$ is the hadronic Mandelstam invariant. 
To make an educated guess of the resummation scale, 
we have studied the partonic configuration which has the largest weight at hadron level;
due to the PDF suppression at large $x$, this happens to be the smallest possible value of $s$.
The choice $s=s_{\text{min}}$ satisfies this requirement but introduces an additional technical problem, namely the presence of soft divergences in the amplitude.
To avoid this issue when computing the curves in figure~\ref{fig:amplcoll:Cpth1} we have set $s=s_{\text{min}}+s_{\text{soft}}$ with $s_{\text{soft}}=(100 $ GeV$)^2$.
We have verified that the results are weakly dependent on the specific value of $s_{\text{soft}}$, as shown in figure~\ref{fig:auxdepcombined} (left plot)
where the bands describe the results, as a function of the Higgs mass, obtained with a variation of $s_{\text{soft}}$ in a range $[1/10,10]$
with respect to the central choice.
In particular we remark that the scale prediction is stable for small values of $s_{\text{soft}}$, i.e. in the soft-emission region, phenomenologically the most relevant.

In figure~\ref{fig:auxdepcombined} (right plot) we show the dependence of the scale determination
on the value assigned to $\bar C$.
The bands describe the results, as a function of the Higgs mass, obtained by varying the parameter in the interval $\bar{C} \in [0.05,0.2]$.
As expected, e.g. from the inspection of figure~\ref{fig:amplcoll:Cpth1},
there is a direct proportionality between the value of $\bar{C}$ and the resulting scale $w$. 

Due to the assumptions used in our procedure, we stress that the determination of the central value for $w$ does not have an absolute meaning.
It is rather the starting point to define an interval of reasonable values for the scale $w$ that in turn should be used to compute an uncertainty
band for the transverse momentum distribution.

 \section{Standard Model phenomenology}
\label{sec:smmass}
We consider now the evaluation of the Higgs transverse momentum distribution in proton-proton collisions at the LHC in the SM.
We use the analytic results of~\cite{deFlorian:2011xf} implemented in the public code \hres~and the shower Monte Carlo implemented in the \powhegbox~\cite{Bagnaschi:2011tu}. 
For the former, 
we study the impact of different choices of the resummation scale $\mu_{\text{res}}$, 
while with the latter 
we vary the value $h$ which enters the damping factor $D_h$. 
In both cases we consider the possibility of a separate treatment of the top and of the bottom quark contributions.
In the numerical analysis 
we use $m_{t} = 172.5$ GeV, $m_{b} = 4.75$ GeV, 
the PDF sets MSTW2008nlo68cl and MSTW2008nnlo68cl~\cite{Martin:2009iq} with their corresponding values of $\alpha_s(m_Z)$. 
We chose $\mu_R = \mu_F = m_H$ as the renormalization and factorization scales. 
We use PYTHIA8~\cite{Sjostrand:2007gs,Sjostrand:2006za} with the tune AU-CT10 to shower the \powheg~events.
This specific tune was chosen since it is the same used by the ATLAS collaboration for their Higgs analyses.
The center of mass energy at the LHC has been assumed to be $\sqrt{S} = 13$ TeV.

\subsection{Comparison of \powheg~and \hres}
\begin{figure}[t]
  \centering
  \includegraphics[width=\textwidth]{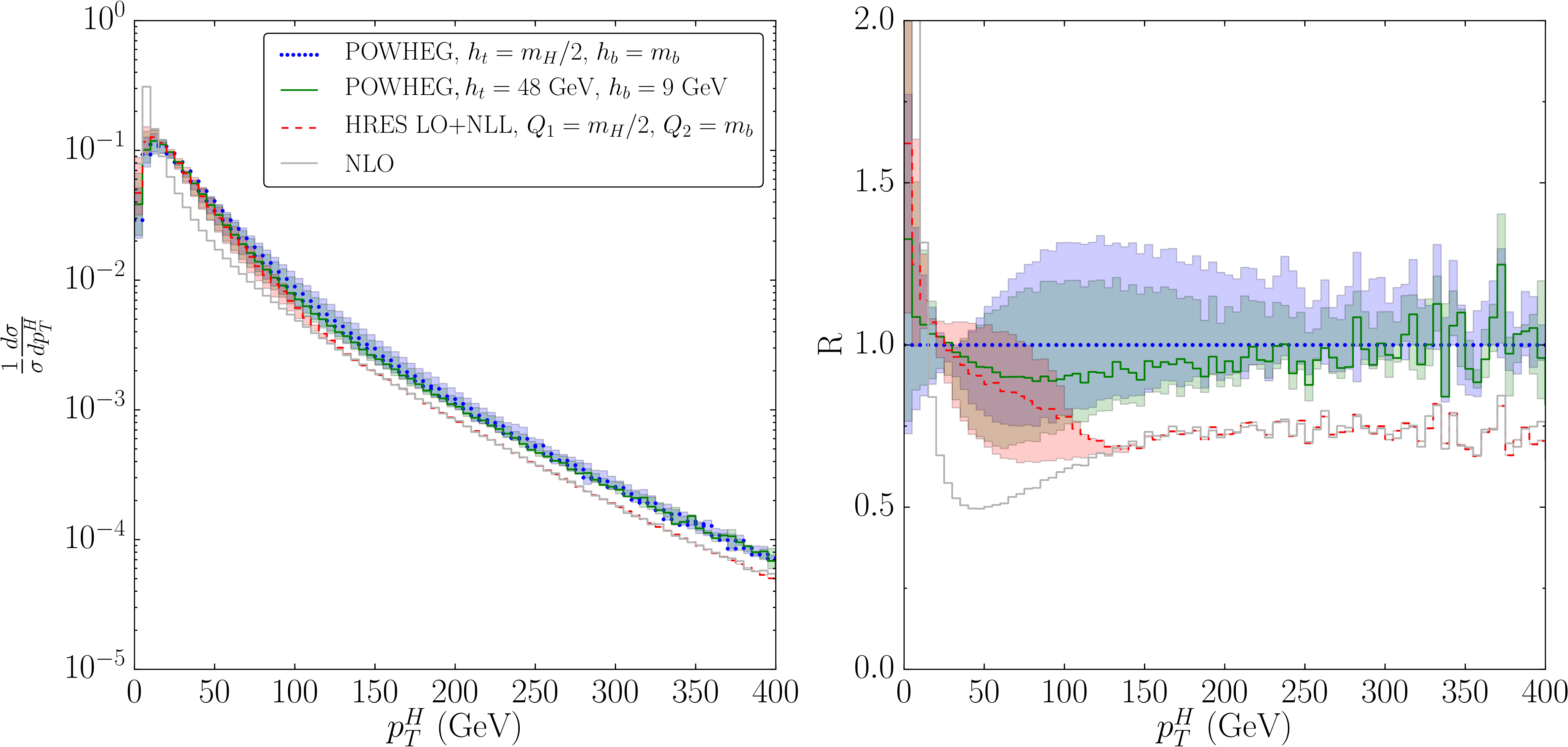}
  \caption{Shape of the transverse momentum distribution for a SM Higgs boson of $m_H = 125$ GeV as computed by \hres~and \powheg~, for different values of the scales.
    On the left we show the absolute value of the shape, while on the right we normalize the results to the one obtained with our $h_t = m_H/2$ and $h_b = m_b$. 
    In dotted blue we show the result obtained with $h_t = m_H/2$ and $h_b = m_b$; with a continuous green we show the prediction obtained with $h_t= w_t = 48$ GeV and $h_b = w_i = 9$ GeV;
    the dashed red line is prediction obtained with HRES at LO+NLL, with $Q_1 = m_H/2$ and $Q_2 = m_b$. For all the three curves we show the corresponding uncertainty bands using the same colors.
    With a continuous gray line we show the results obtained at NLO.
}
  \label{fig:hresvspowheg}
\end{figure}

\begin{figure}[t]
  \includegraphics[width=\textwidth]{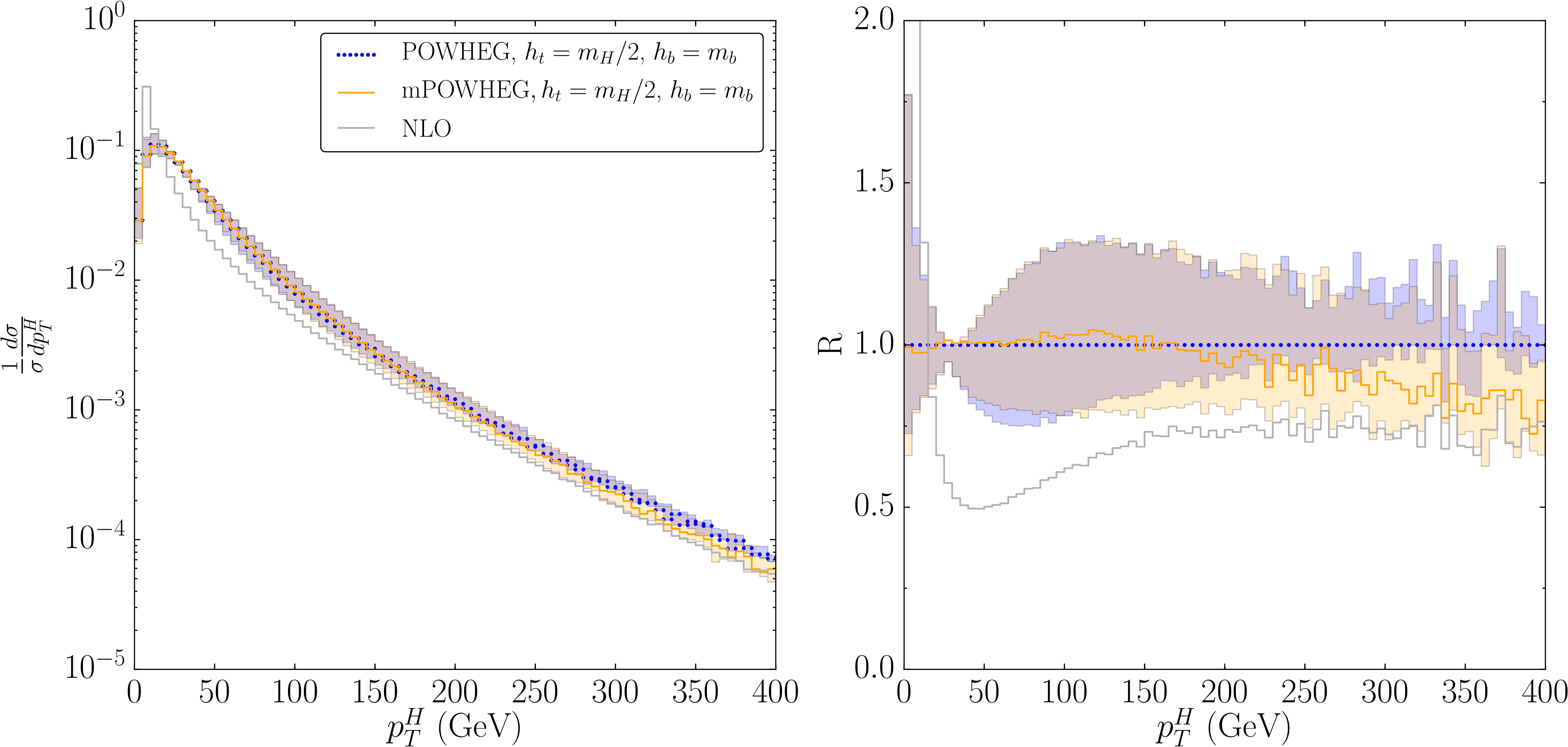}
  \caption{Comparison of the $\pth$ spectrum of the Higgs boson in the SM, with our scale choices, using the default \powhegbox~implementation (blue) and the one with the modified
           {\tt SCALUP} prescription for the remnant events (orange).}
         \label{fig:scalup}
\end{figure}

The gluon fusion process, including the top and the bottom quark diagrams in the scattering amplitude, is a three-scale problem, 
as was already stressed in ref. \cite{Grazzini:2013mca} and as we have seen in the previous sections: the Higgs mass, the value of $\pth$ and the mass of the quark.
The bottom quark contributions spoil the validity of the factorization hypothesis for $\pth$ values smaller than in the top quark case and require a dedicated treatment.
In order to make explicit the role of the top and of the bottom quarks,
the squared matrix elements can be rearranged as
\begin{align}
  \left|\mathampl(\text{top}+\text{bot})\right|^2 = \left|\mathampl(\text{top})\right|^2 + \left[ \left|\mathampl(\text{top}+\text{bot})\right|^2 - \left|\mathampl(\text{top})\right|^2 \right]\,,
  \label{eq:smpheno:scalessplit}
\end{align}
where we have put in round bracket the quarks that run in the loops of the diagrams. 
The square brackets contain the top-bottom interference terms and the square of the modulus of the bottom amplitude.
The rationale behind this rearrangement is that in the SM the dominant contribution to the gluon fusion is due to the top quark diagrams, while the bottom quark diagrams yield a correction to the former; 
it is thus reasonable to make one dedicated scale choice for the top quark
and a second scale choice for all the other terms, even if they still include
top quark diagrams via interference terms.
We recall that by construction the total cross section does not depend on the value of the resummation scale in \hres~ (or equivalently of the scale $h$ in \powheg). 
This fact allows us to write the following identity
\begin{align}
  \sigma(\text{top}+\text{bot}) = \sigma(\text{top},\mu_t) + \left[ \sigma(\text{top}+\text{bot},\mu_b) - \sigma(\text{top}, \mu_b ) \right]\,,
  \label{eq:smpheno:threexsecs}
\end{align}
where here and after, with a slight abuse of notation, we have introduced the symbol $\sigma(q,\mu)$ to indicate the total cross section evaluated with the quark $q$ in the loops, 
using, in the numerical code, the matching parameter at the scale $\mu$. The latter is the resummation scale $Q_i$ in \hres~and the scale $h$ in \powheg.

This equation is trivial for the total cross section, and represents a possible recipe for the evaluation of
differential observables, specifically the Higgs boson transverse momentum.\footnote{In \powheg, at the differential level, the extraction of a specific contribution by subtraction is bound to introduce spurious terms
  due to the fact that the Sudakov form factor is non-universal. 
However, due to our specific scale choices that guarantee 
a good accuracy of the collinear approximation
in the $\pth$ range where
the Sudakov form factor has its major effect,
we can argue that in this region the argument in the exponent of the Sudakov factor is well approximated by the relevant universal expression
$R/B \simeq \alpha_s P_{ij}/t$,
limiting the impact of the spurious terms.}

For our phenomenological analysis we use two scales, one for the squared matrix element with only the top quark and one for the other contributions, to allow a comparison with the results presented in
ref.~\cite{Grazzini:2013mca}; we use a combination analogous to the one of eq.~(\ref{eq:smpheno:threexsecs}) to evaluate also the differential distributions.

In section~\ref{sub:depaux} we have given an estimation of the uncertainty in the determination of the scales $w$
by varying the auxiliary parameters that we have used in our computation. In theory it is possible to use the range of scales obtained
with such a procedure as the range of values to be used for the matching parameter to estimate the uncertainty on the prediction for the transverse momentum distribution.
However these values depend in a non-trivial manner on the Higgs mass considered.
We observe that a variation by a factor of $2$ of the central value
widely covers the range of scales that we find with our explicit computation,
thus yielding a conservative assessment of the uncertainty.
To simplify the uncertainty-estimation procedure we have then decided to compute the uncertainty bands using the following standard prescription:
we consider the 9 combinations of the pairs $(\mu_t,\mu_b)$ of the two matching parameters, 
which can be obtained from the sets $(\bar\mu_t/2,\bar\mu_t,2 \bar\mu_t)$ and
$(\bar\mu_b/2,\bar\mu_b,2 \bar\mu_b)$, where
we called $\bar\mu_t$ and $\bar\mu_b$ the respective central values, and we take the envelope of all the predictions.

We consider the three following cases
and, in each of them, we compute the uncertainty band according to the rule described above:
\begin{enumerate}
\item
we use \powheg~and we set the scale of the top quark diagrams $h_t=\mh/2$ and the scale of the bottom quark contributions $h_b=m_b$
\item
we use \powheg~and follow the analysis described in section \ref{sec:hggcoll}
and in particular the values of table \ref{tab:hthbscalar}:
we set $h_t = w_t=48$ GeV and $h_b = w_i=9$ GeV. The $w_i$ is chosen over $w_b$ 
since the interference terms yield a larger contribution to the process than the bottom quark squared matrix elements.
\item
we use \hres~at LO+NLL accuracy and set the resummation scale of the top quark diagrams $Q_1=\mh/2$ and the resummation scale of the bottom quark contributions $Q_2=m_b$, following the choices of ref.\cite{Grazzini:2013mca};
\end{enumerate}

The distributions obtained with \hres~and \powheg~share the same 
matrix elements that describe at NLO-QCD the inclusive Higgs boson production,
and differ by subleading NNLO and by higher order terms, 
which might nevertheless be numerically relevant.

The comparison of the shape\footnote{With the term shape we mean that we have normalized the differential distribution to $1$.}
of the $\pth$ distribution, in figure~\ref{fig:hresvspowheg}, of the results of item 1 (blue dot line) and 3 (dashed red line)
is meant to expose the differences of the two codes taken with their default setup,
when they are run with the same accuracy for the total cross section, NLO-QCD,
and with the same value for the matching parameters. On the left we show the absolute comparison of the results, while on the right
we show the ratio of the different predictions over the one obtained with \powheg~and the \hres~scale choice (item 1).

As discussed in section \ref{sec:generalresumm}
the two basic formulae used to generate the Higgs $\pth$ spectrum differ by subleading ${\cal O}(\alpha_s^2)$ and higher-order terms, part of which are controlled by the resummation scale in \hres~or by the $h$ scale in \powheg.
For the above reason, even if we assign the same numerical values to the scales $Q$ and $h$, we expect a certain level of discrepancy for the central predictions.

Indeed we see that in the region where resummation effects are relevant, the two codes behave differently, with \hres~giving a softer distribution than \powheg.
Specifically, the shape of the distribution produced by \hres~is larger  than the one from \powheg~for $\pth\le 50$ GeV, while for higher $\pth$ the behavior is the opposite.
In the high-$\pth$ region, for $\pth\ge\mh$, we see that the \hres~result coincides with the fixed-order distribution: 
in fact, the code \hres~uses the full matched expression for $\pth$ values smaller than $\mh$ and implements a smooth transition to
the pure fixed-order expression, which is used in the high-$\pth$ tail;
for this same reason, the \hres~resummation scale uncertainty band vanishes in this part of the spectrum.

In the high-$\pth$ range \powheg~shows a distribution harder than the fixed-order one,
because of the showering effects applied on top of the \powheg~ formula for the first emission.

Since \hres~does not include non-perturbative effects, which are present in the selected tune of the {\tt PYTHIA} shower,
an additional problem in the comparison emerges:
the non-perturbative effects are
relevant at small transverse momenta of the radiated partons. 
In addition, in the low-$\pth$ region, the different expression of the \hres{} and \powheg{} Sudakov form factors (for the latter see equation \ref{eq:sudakov}) has a role to determine the precise shape of the distribution.
By construction, the unitarity constraint, 
that forces the total cross section to be always preserved,
implies an anti-correlation between the low-$\pth$ and the high-$\pth$ parts of the spectrum. 

The comparison in figure~\ref{fig:hresvspowheg} of the results of approximations 1 and 2
shows the sensitivity, within the \powheg~formulation of the matching, 
to the $h$ scale variations.
The two central values lie in the uncertainty bands obtained with the other scale choice. 
The main difference can be observed at small $\pth$, whereas the deviation
for $50\le\pth\le 150$ GeV can be interpreted as a consequence of the unitarity constraint.

We observe, by using $h_t = w_t$ and $h_b = w_i$, an accidental improvement of the agreement between \hres~and \powheg~
in the region of $\pth < 100$ GeV, where the two central values lie close to each other.

In figure~\ref{fig:scalup} we present the impact in \powheg~ of a different choice of the variable {\tt SCALUP}, as discussed in section \ref{sec:scalup}.
We set {\tt SCALUP}=$h_t$, a constant value,
while we keep unchanged all the other parameters
and in particular the value of the scales $h_{t,b}$ in the damping factor $D(h)$.
The choice for the {\tt SCALUP} value is
in accordance with the dominant role played by the top-quark loop in the SM.
We observe that the central prediction of this modified \powheg~ version
is lower than the default one for $\pth\ge 200$GeV
and tends to recover the fixed-order distribution at large transverse momenta.
We interpret the reduction of the differential cross section at large $\pth$ as due to the missing contribution in this region from the PS emissions.
The accuracy of the latter is questionable, since the PS is based on the soft/collinear approximation and might be inadequate to describe large-$\pth$ radiation.


%
 \section{Beyond SM phenomenology}
\label{sec:bottdom}

The description of the Higgs transverse momentum distribution in the SM,
with $\mh=125$ GeV,
is characterized by the dominant role played by the top-quark contribution, such that the bottom-quark effects can be treated as a correction.
Moreover, with a light scalar Higgs, the HQEFT limit is a good approximation of the full SM, and the determination of the scale of validity of the collinear approximation (and hence of the applicability range of the resummation techniques)
reduces to a problem involving only $\mh$ and $\pth$.

At variance with the previous case, and still in the SM, 
we know that with a heavy Higgs boson, the description of the $\pth$ distribution is a multiscale problem; indeed, the minimal energy scale necessary
to produce the final state immediately probes the top-quark loop.

In a generic BSM scenario it is possible to consider
enhanced couplings of the bottom quarks to a relatively heavy Higgs boson, scalar or pseudoscalar.
In these configurations, our intuition, 
accustomed to a light SM-like Higgs phenomenology, 
may fail in the determination of the correct regime where the resummation techniques can be safely applied.
Since a priori we do not know exactly how the contributions from the different quarks interplay in the full result, following ref.~\cite{Harlander:2014uea}, we can generalize eq.~(\ref{eq:smpheno:threexsecs}) to
\begin{align}
  \sigma(\text{top}+\text{bot})&= \sigma(\text{top},\mu_t) + \sigma(\text{bot},\mu_b) \nonumber \\
&+  \left[ \sigma(\text{top}+\text{bot},\mu_i) - \sigma(\text{top},\mu_i) - \sigma(\text{bot}, \mu_i) \right]\,,
  \label{eq:mssmpheno:threescales}
\end{align}
where the last term allows us to use a separate scale for the top-bottom interference term.
As before, the parton level analyses discussed in section \ref{sec:hggcoll} provide a model independent {\it ansatz} for the three relevant scales, $\mu_{t,b,i}$:
these are the scales $w_t$,$w_b$ and $w_i$, listed in tables \ref{tab:hthbscalar} and \ref{tab:hthbpseudoscalar}  as a function of the Higgs boson mass.

In order to illustrate the phenomenological consequences of our study, we show our predictions in the 2HDM and in the MSSM\footnote{
A detailed comparison with the approach of ref.~\cite{Harlander:2014uea}
is currently ongoing \cite{Bagnaschi:2015bop}.
} and we compute the uncertainty bands with an extension of the procedure described in section \ref{sec:smmass}: 
we consider all the 27 combinations of the three matching scales and then take their envelope.
The range of scales spanned represents again a conservative choice to assess the matching uncertainty.

\subsection{2HDM phenomenology}

\begin{figure}[t]
\begin{center}    
  \includegraphics[width=\textwidth]{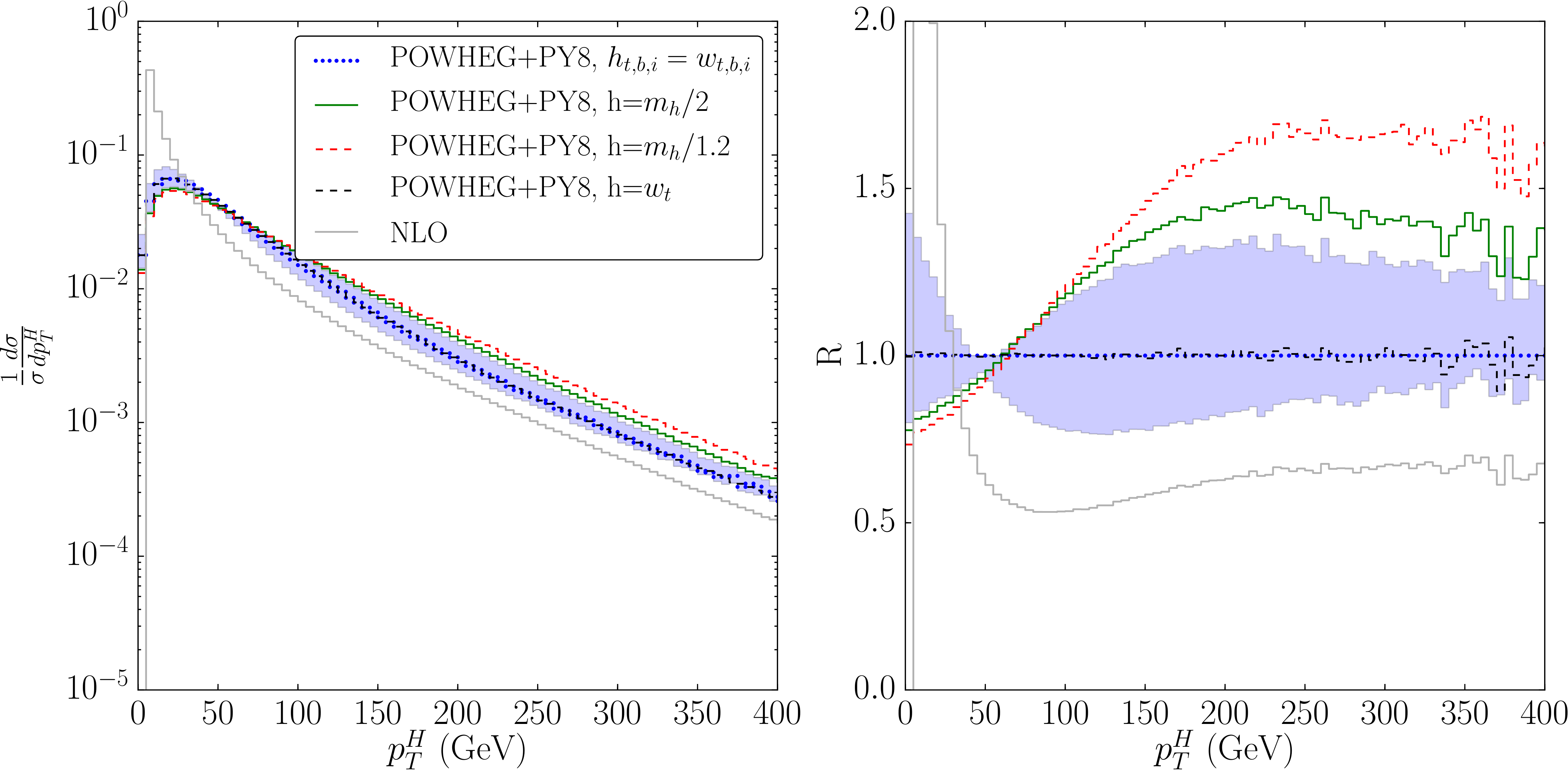}\\
  \caption{Shape of the transverse momentum distribution for the heavy CP-even scalar H, as computed by the {\tt gg\_H\_2HDM} generator in the 2HDM scenario A. 
On the left we show the absolute value of the shape, while on the right we normalize the results to the one obtained with the scales determined by our procedure, $h_t = w_t = 96$ GeV,$h_b = w_b = 58$ GeV and $h_i = w_i = 17$ GeV.
In dotted blue we show the result obtained with our scale choice, its uncertainty band drawn in lighter blue; with a continuous green (dashed red line) we show the prediction obtained with $h=m_h/2$ ($h=m_h/1.2$). In dashed black we show the results obtained with a single run with the scale $h$ set to $w_t$. Finally in gray we show the NLO prediction.
\label{fig:powhegpt2hdmA} }
\end{center}
\end{figure}

\begin{figure}
\begin{center}    
  \includegraphics[width=\textwidth]{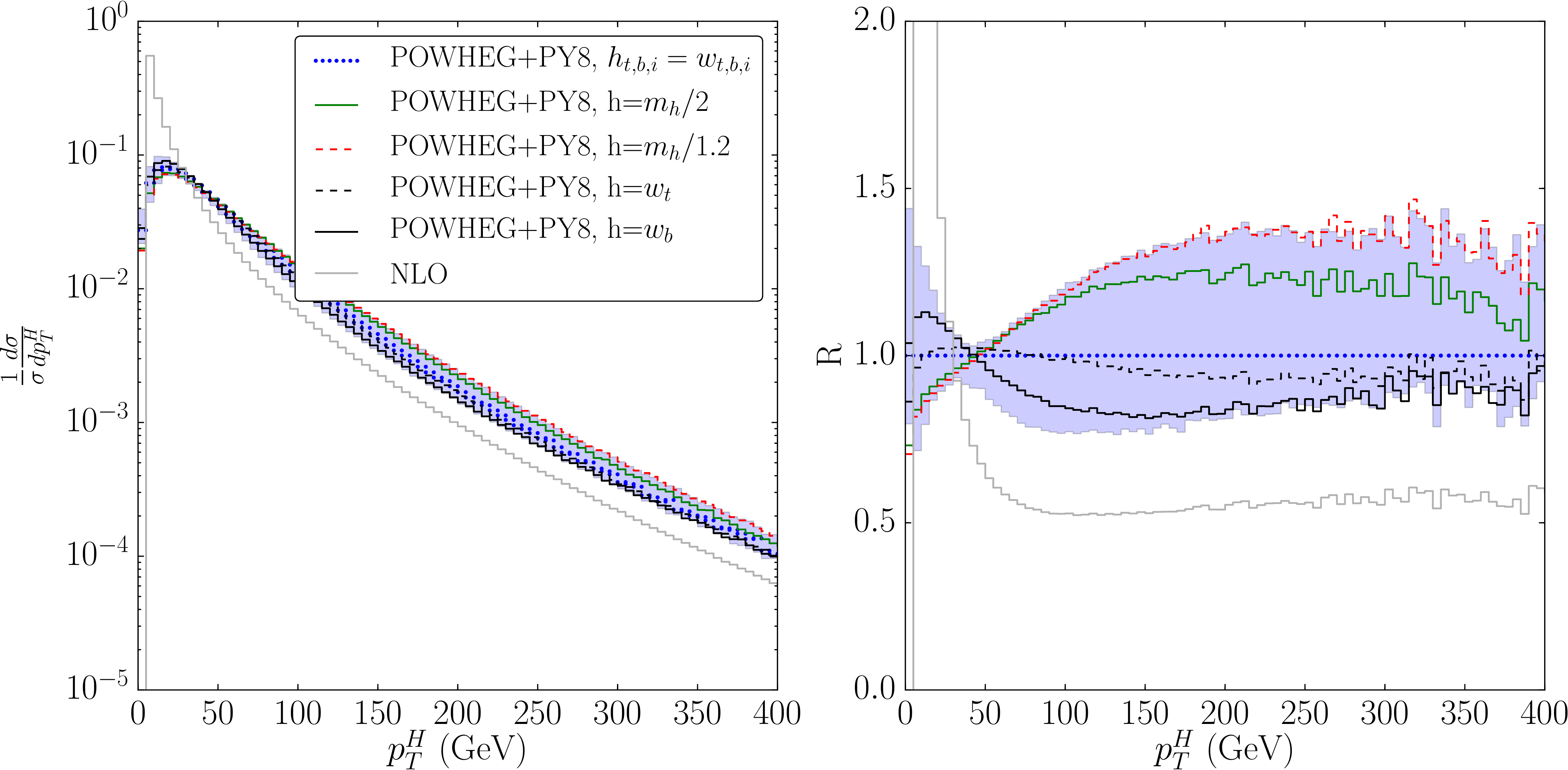}\\
  \caption{Same as in figure \ref{fig:powhegpt2hdmA} but for 2HDM scenario B. Here the dashed black line is obtained with a single run with $h = w_t$ while the continuous black line corresponds to $h = w_b$.
\label{fig:powhegpt2hdmB} }
\end{center}
\end{figure}

\begin{figure}
\begin{center}    
  \includegraphics[width=\textwidth]{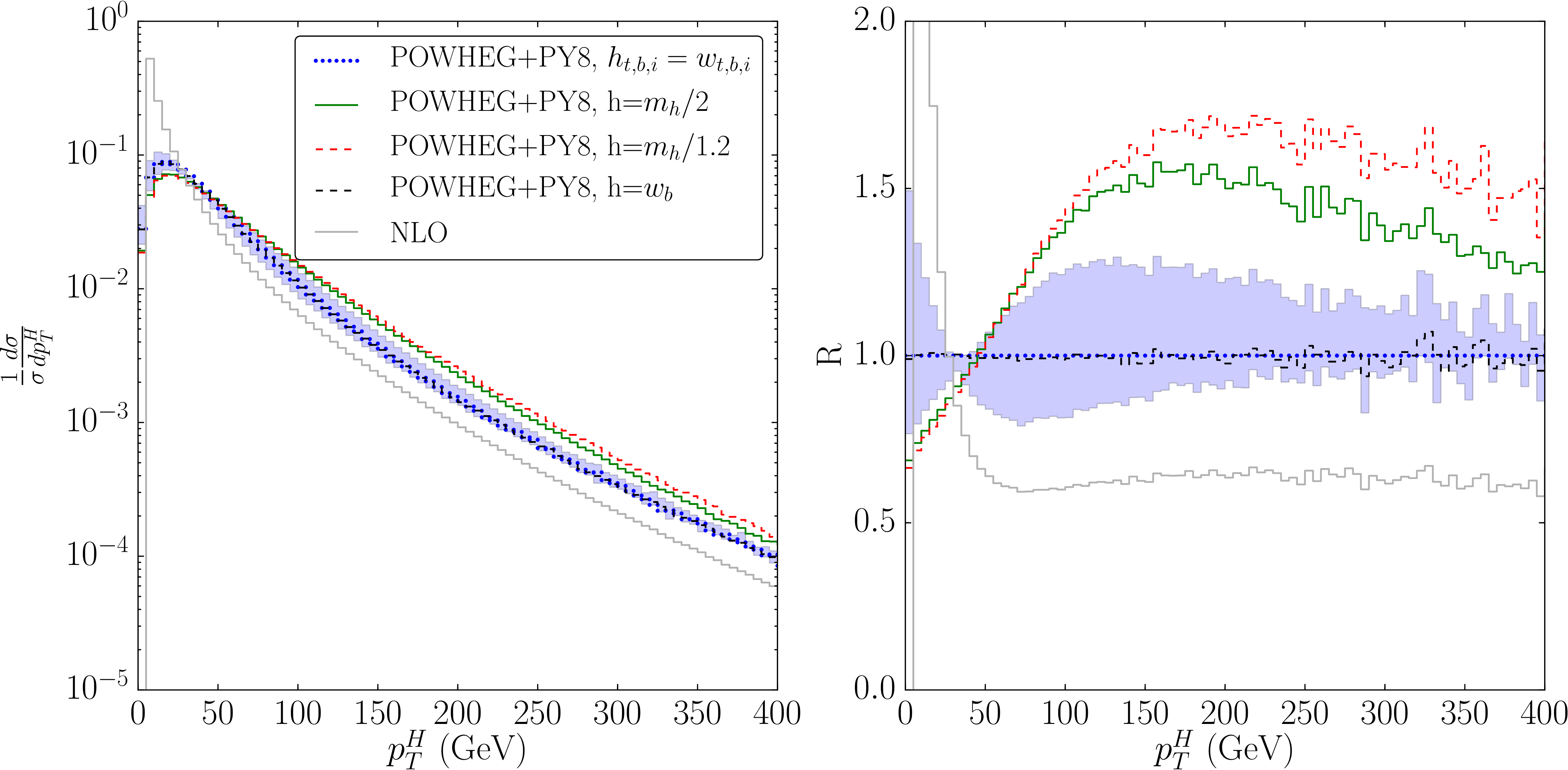}\\
  \caption{Same as in figure \ref{fig:powhegpt2hdmA} but for 2HDM scenario C. Here the dashed black line is obtained with a single run with $h = w_b$.
\label{fig:powhegpt2hdmC} }
\end{center}
\end{figure}

\begin{table}[!h]
\centering
\begin{center}
\begin{tabular}{@{}c|ccccc}
\toprule
\multicolumn{6}{c}{\bf 2HDM scenarios}\\
\midrule
 Parameter  & Scenario A & & Scenario B & & Scenario C \\
\cmidrule{1-1} \cmidrule{2-2} \cmidrule{4-4} \cmidrule{6-6}
$m_H$ (GeV) & $500$  & & $500$ & & $500$ \\
$\tan\beta$ & $1$ & & $12$ & & $50$ \\
$\sin(\beta-\alpha)$ & $1$ & & $1$ & & $1$ \\
\bottomrule
\end{tabular}
\end{center}
\caption{Values of the relevant 2HDM parameters for the three type-II scenarios considered in the text.}
\label{table:2hdm}
\end{table}

We consider the type-II 2HDM. We adopt a purely heuristic approach to show the impact of our study, choosing the 2HDM parameters that are relevant for the gluon fusion process
by following only the requirement that they represent three different scenarios: one where the cross section is dominated by the top-quark;
one where the contribution of the top and the bottom quark are of the same order of magnitude; and one where the process is dominated by
the bottom quark matrix elements. The explicit values for the parameters are reported in table~\ref{table:2hdm} for all the three scenarios.
In all three cases we choose to study a heavy Higgs of $m_H = 500$ GeV. The corresponding values for the scales are ${w_t=96}$ GeV, $w_b=58$ GeV and $w_i=17$ GeV.
For the simulation we adopt the Monte Carlo generator
{\tt gg\_H\_2HDM} available in the \powhegbox.


We now present our best predictions obtained with the three-scale procedure and check how well they are approximated by a one-scale approach.

In fig~\ref{fig:powhegpt2hdmA} we show the results for the first scenario. In this case we have that the process is dominated by top quark contribution.
Indeed we notice that the three scales result is well approximated by the one scale result with the scale taken equal to the top scale.

On the other hand, in the second case shown in figure \ref{fig:powhegpt2hdmB}, we have that the contributions coming from the two quarks are of the same order of magnitude.
In this case we observe that the result obtained by using three scales is not recovered by simulations with just a single scale, with either the top or the bottom one.

Finally, in figure \ref{fig:powhegpt2hdmC} we see that in the bottom dominated scenario, we have a similar situation as in the top dominated case, though the scale to be used in
a one scale run is $w_b$ instead of $w_t$.

In all three cases we stress that the using values of the order of $m_H/2$ or $m_H/1.2$ for the matching parameter $h$ yields results that
are in the best case at the limit of the uncertainty band.

\subsection{MSSM phenomenology}

\begin{figure}
\begin{center}    \includegraphics[width=\textwidth]{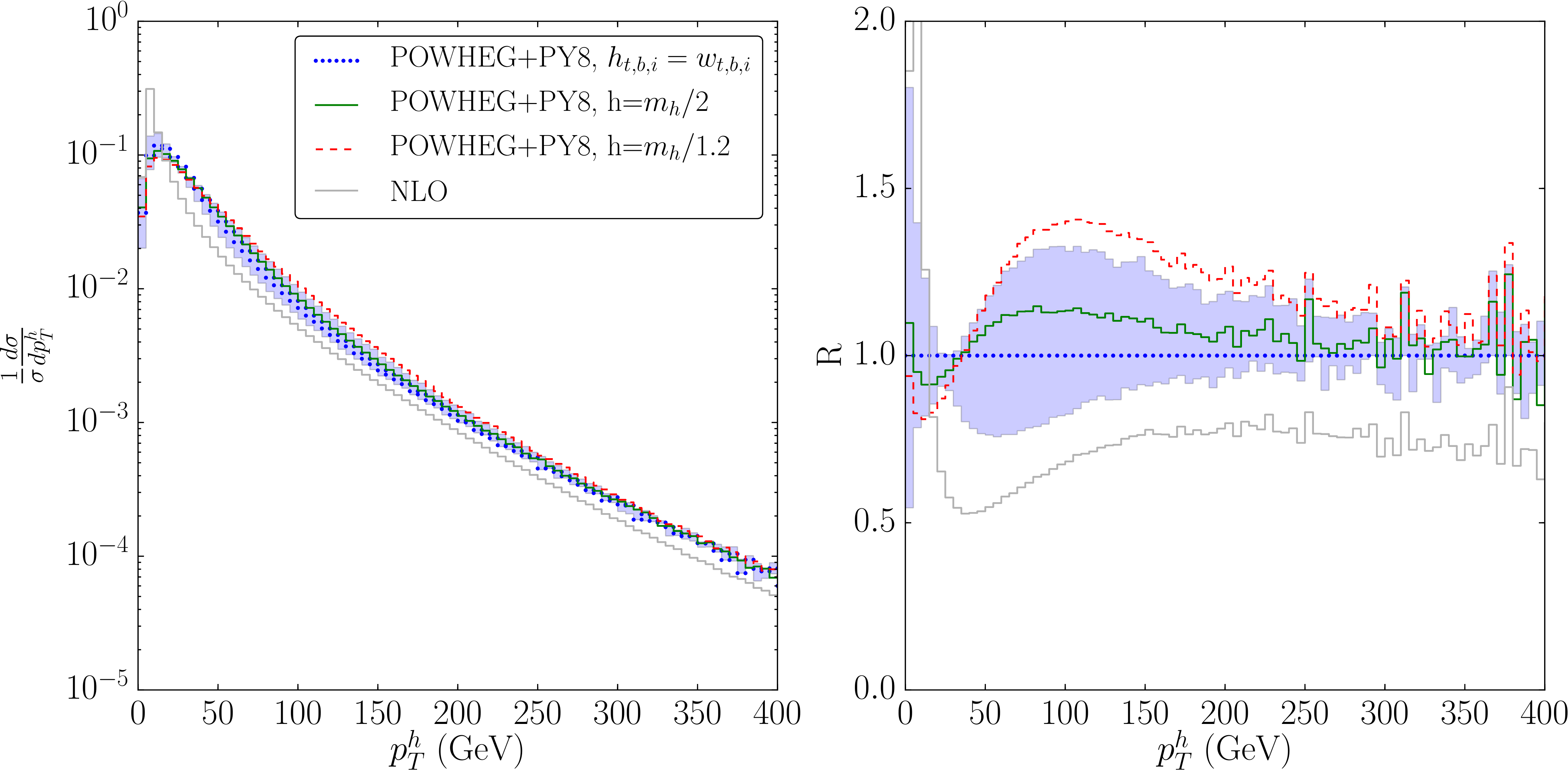}
  \caption{Shape of the transverse momentum distribution for light CP-even scalar h, as computed by the {\tt gg\_H\_MSSM} generator for $\tan\beta=17$ and $m_A=500$ GeV in the $m_h^{\text{mod}+}$ scenario. 
On the left we show the absolute value of the shape, while on the right we normalize the results to the one obtained with our scale choice. In dotted blue we show
the result obtained with our scale choice, its uncertainty band drawn in lighter blue; with a continuous green (dashed red line, we show) the prediction obtained with $h=m_h/2$ ($h=m_h/1.2$).
\label{fig:powhegptmssm1} }
\end{center}
\end{figure}

\begin{figure}
\begin{center}  
\includegraphics[width=\textwidth]{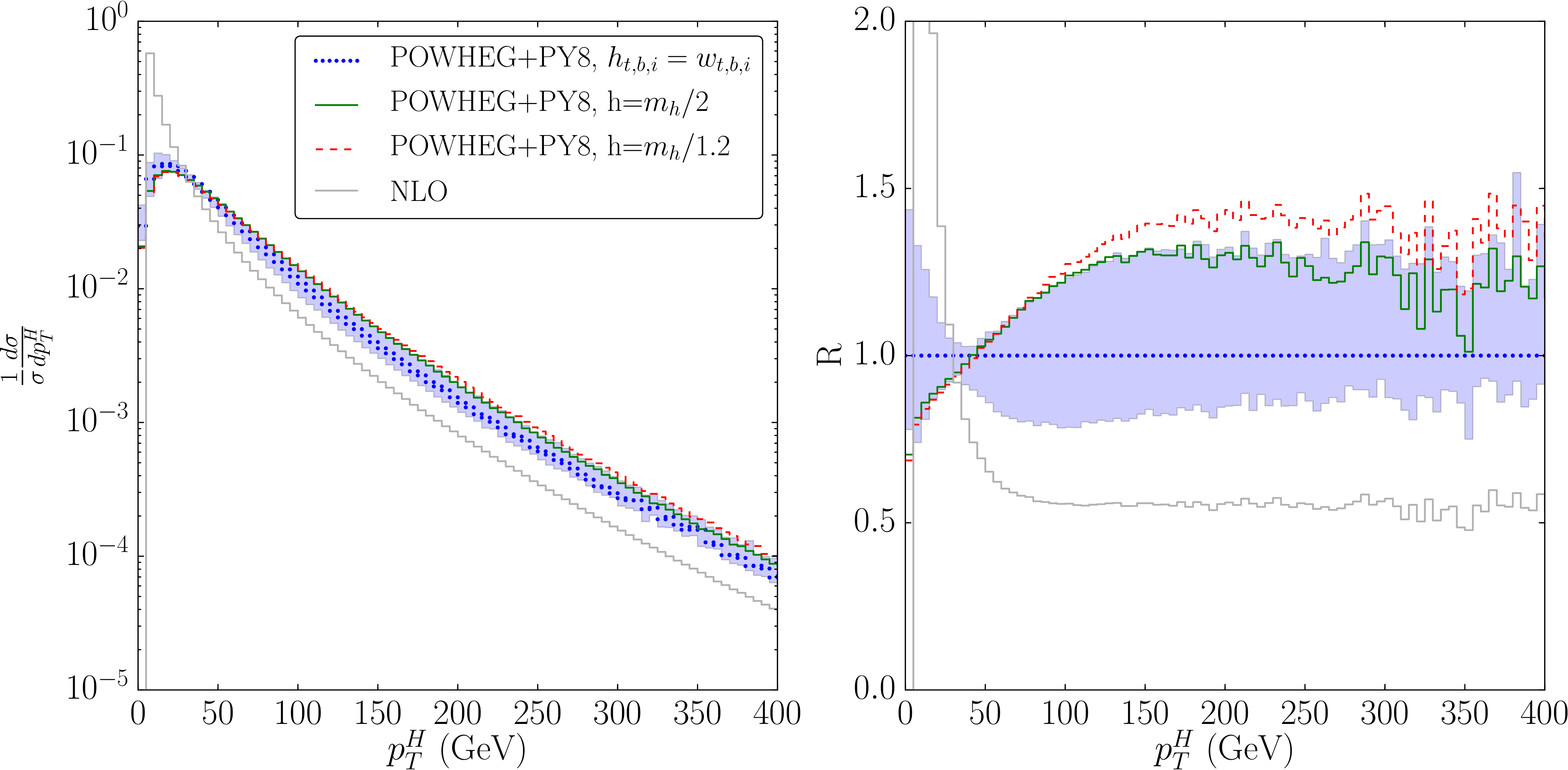}
\caption{Same as in figure~\ref{fig:powhegptmssm1}, but now for the heavy CP-even scalar H. 
\label{fig:powhegptmssm2}   }
\end{center}
\end{figure}

We consider an explicit example in the MSSM by taking, in its parameter space, 
a point still allowed by the most recent available data,
according to the analysis of ref.~\cite{Carena:2013qia},
and to the results of the code {\tt HiggsBounds}
\cite{Bechtle:2008jh,Bechtle:2011sb,Bechtle:2013gu,Bechtle:2013wla}.
The same point has been considered also in ref.\cite{Harlander:2014uea}.
We choose the so called $m_h^{\textrm{mod}+}$ scenario defined in ref.~\cite{Carena:2013qia}
and set $M_A=500$ GeV and $\tan\beta=17$ to fully specify our input parameters; as a result we obtain that the masses of the two CP-even Higgses are respectively
$m_h=125.6$ GeV and $m_H=499.9$ GeV.
The corresponding values of the $w$ scales are: ${w_t=96~(109)}$ GeV, $w_b=58~(58)$ GeV and $w_i=17~(14)$ GeV, for a scalar (pseudoscalar) boson.
We use these values to set the $\mu_{t,b,i}$ parameters that enter eq.~(\ref{eq:mssmpheno:threescales}). 
For the simulation we adopt the Monte Carlo generator
{\tt gg\_H\_MSSM} available in the \powhegbox.
In the simulation we include the full particle content of the MSSM.
We do not expect an important contribution from the squarks
because in this point of the MSSM parameter space their masses are, respectively,
$m_{\tilde{t}_1} = 876$ GeV, $m_{\tilde{t}_2} = 1134$ GeV, $m_{\tilde{b}_1} =  1007$ GeV and $m_{\tilde{b}_2} = 999$ GeV.

With this specific parameter choice,
the light CP-even Higgs is similar to the SM scalar, 
not only for the total cross section, 
but also for the shape of the $\pth$ distribution.
The heavy CP-even scalar and the pseudoscalar bosons have instead different
properties, because of the different coupling strength to the top and to the bottom quarks.

In figures~\ref{fig:powhegptmssm1} and~\ref{fig:powhegptmssm2} we show the results for the shape of the transverse momentum distribution in the case of the light CP-even Higgs (top plot) and of the heavy CP-even Higgs (bottom plot). 
We do not show the plot for the pseudoscalar since we expect a behavior similar to the one of the heavy Higgs. 
Besides plotting the central values and the uncertainty band corresponding to our scale choice, 
we also show the results obtained  with only one matching scale, 
with the commonly used prescriptions $h = m_{h,H}/2$ and $h = m_{h,H}/1.2$. 
We observe that the three choices yield a different shape of the distribution in the soft region where resummation effects are important: 
the scale choices $h = m_{h,H}/2$ and $h = m_{h,H}/1.2$
give a suppression in the first bins and an enhancement for $\pth$ larger than $40$ GeV with respect to the distribution
obtained following eq.~(\ref{eq:mssmpheno:threescales}). 
In the case of a light Higgs, we see that the central value obtained with $h=m_h/2$ is contained in the uncertainty band of the prediction computed by using three scales,
while the result corresponding to $h=m_h/1.2$ is at the edge of the same uncertainty band.
In the case of the heavy CP-even Higgs, 
the $h = m_{H}/2$ and $h = m_{H}/1.2$ curves 
lie outside the uncertainty band of the three-scale result;
they deviate from its central value
by ${\cal O}(40\%)$, both in the low- and in the high-$\pth$ tails.


%
 \section{Conclusions}

The study of the Higgs transverse momentum distribution may provide important insights about the properties of the recently discovered scalar resonance.
The theoretical prediction of this observable requires, in the region of small $\pth$ values, the resummation to all orders of terms enhanced by powers of
$\log(\pth/\mh)$, while at large values of $\pth$, fixed-order calculations provide the most accurate description available.
The consistent matching of the two approaches requires the introduction of a momentum scale,
that separates the soft and the hard $\pth$ regions.

Since the validity of the resummation formalism relies on the collinear factorization of the squared matrix elements describing real parton emissions,
we investigated the accuracy of the collinear approximation in the gluon fusion process, in the presence of an exact description of the top and bottom quarks running in the virtual loop.
The discussion involves three scales, namely the Higgs mass, the Higgs transverse momentum and the quark masses.

Relying on the collinear singularities structure of the $\mathcal{O}(\alpha_s)$ real matrix elements,
we determined, in a model independent way, 
as a function only of the Higgs and the quark masses,
three scales, $w_t, w_b$ and $w_i$,
associated to the terms in the full squared matrix elements containing only the top-, only the bottom-quark contributions or the top-bottom interference terms.
Their values, presented in tables \ref{tab:hthbscalar} and \ref{tab:hthbpseudoscalar}
and, with a finer scan of the Higgs mass, in appendix \ref{app:hthbscan},
represent our main result.
These scales are derived from a parton-level analysis and can be eventually used in any hadron-level computation (analytic or Monte Carlo) of the Higgs $\pth$ distribution,
following eq. (\ref{eq:mssmpheno:threescales}).
They indicate the upper limit of the $\pth$ range where the resummed part of the cross section can be evaluated in a reliable way,
because of the good accuracy of the collinear approximation of the full squared matrix elements.
They represent an {\it ansatz} for the matching scales,
whose values do not have an absolute meaning, but are rather the starting points to build an uncertainty band.

The procedure to compute an uncertainty band is described in section \ref{sec:smmass} and offers a simple but quite conservative recipe
to derive this band.
A more aggressive approach would exploit the scales obtained with a variation of the parameter $\bar C \in [0.05,0.2]$, as discussed in section \ref{sec:hggcoll}.

Our analysis is relevant for an accurate prediction of the Higgs $\pth$
distribution, both in the SM and in BSM scenarios.
In the latter case,
our approach allows us to decompose
the different contributions to the $\pth$ distribution,
also in the presence of a non trivial interplay between
the Higgs transverse momentum and 
the Higgs, top and bottom masses, 
for any generic ratio between the strength of the couplings of the Higgs boson to the top and to the bottom quarks.

The description of the Higgs transverse momentum distribution, based on the use of three different scales
for the matching parameter, represents our best {\it ansatz} for this observable.
We remark, however, that in various cases this result can be accurately approximated with only one run that uses one single scale, the one associated to the dominant contribution to the scattering amplitude.
This conclusion is obviously possible only {\it a posteriori}.

We stress the impact of the matching scale determination with one final comment,
relevant in the context of the searches for new heavy scalars,
referring to the results shown in figures~\ref{fig:powhegptmssm1} and \ref{fig:powhegptmssm2}.
Our procedure defines the scales $w_{t,b,i}$, 
whose variation in a given range
is then exploited to compute an uncertainty band of the distribution. 
The results presented in section \ref{sec:bottdom} are obtained 
with a conservative choice for the range of scale variation,
as described at the beginning of the same section.
The use of a single-scale simulation,
with the matching scale set equal to the commonly adopted SM value $\mh/2$,
can lead to predictions 
that lie outside of this most conservative uncertainty band described above
and 
that can differ with respect to our best central value by 30-40\%
both in the low- and in the high-$\pth$ tails of the distribution.

\section*{Acknowledgments}

We would like to thank Giancarlo Ferrera, Stefano Frixione, Massimiliano Grazzini and Carlo Oleari for many discussions on the subject.
We thank Giuseppe Degrassi, Robert Harlander, Hendrik Mantler, Pietro Slavich and Marius Wiesemann for carefully reading the manuscript.
We are indebted with Paolo Nason for several suggestions and clarifications about {\tt POWHEG}.

E.~B.~was partially supported by the EU ITN grant LHCPhenoNet, PITN-GA-2010-264564.
A.~V.~ is supported in part by an Italian PRIN2010 grant, by a European Investment Bank EIBURS grant, and by the European Commission through the HiggsTools Initial Training Network PITN-GA-2012-316704.
A.~V.~ thanks 
the LPTHE, the University ``Pierre et Marie Curie'' Paris VI and
the Institut Lagrange in Paris 
for financial support in summer 2013.

\appendix

\clearpage
\section{Scan over the Higgs mass of the scales \texorpdfstring{$w_{t,b,i}$}{wt,wb,wi}}

In the appendix we include two tables with the values of the combined \emph{gg-qg} collinear-deviation scales,
for scalar and pseudoscalar masses between $100$ GeV and $500$ GeV, separately for the top, the bottom
and the interference contribution. The top pole mass has been set to $172.5$ GeV, while the bottom
pole mass is equal $4.75$ GeV, following the prescription by the Higgs Cross Section Working Group (HXSWG).

The merging of the scales was implemented by using the information
on the relative importance of the two partonic subprocess as given by the code {\tt SusHi}\cite{Harlander:2012pb}.
The latter was run at $\sqrt{S} = 13$ TeV using the {\tt MSTW2008nlo68cl} PDF set and setting $\mu_r = \mu_f = m_H$.

\label{app:hthbscan}
\begin{table}[!h]
\centering
\begin{center}
\begin{tabular}{@{}cccccc|ccccc}
\toprule
\multicolumn{11}{c}{\bf Scalar and pseudoscalar collinear deviation scale $w$ (GeV)}\\
\midrule
$\mh$ (GeV) & & $w^{gg+qg}_t$ & $w^{gg+qg}_b$ & $w^{gg+qg}_i$ & & $m_A$ (GeV) & & $w^{gg+qg}_t$ & $w^{gg+qg}_b$ & $w^{gg+qg}_i$ \\
\cmidrule{1-1} \cmidrule{3-5} \cmidrule{7-7} \cmidrule{9-11} 
$100$ &  & $42$ & $14$ & $7$ &  & $100$ &  & $43$ & $14$ & $8$ \\ 
$110$ &  & $44$ & $16$ & $8$ &  & $110$ &  & $46$ & $16$ & $9$ \\ 
$120$ &  & $47$ & $17$ & $8$ &  & $120$ &  & $50$ & $17$ & $10$ \\ 
$125$ &  & $48$ & $18$ & $9$ &  & $125$ &  & $52$ & $18$ & $10$ \\ 
$130$ &  & $50$ & $18$ & $9$ &  & $130$ &  & $53$ & $18$ & $10$ \\ 
$140$ &  & $52$ & $19$ & $10$ &  & $140$ &  & $57$ & $19$ & $11$ \\ 
$150$ &  & $55$ & $21$ & $11$ &  & $150$ &  & $61$ & $21$ & $12$ \\ 
$160$ &  & $58$ & $22$ & $11$ &  & $160$ &  & $65$ & $22$ & $13$ \\ 
$170$ &  & $61$ & $23$ & $12$ &  & $170$ &  & $70$ & $23$ & $14$ \\ 
$180$ &  & $64$ & $24$ & $13$ &  & $180$ &  & $75$ & $24$ & $14$ \\ 
$190$ &  & $67$ & $26$ & $14$ &  & $190$ &  & $82$ & $26$ & $15$ \\ 
$200$ &  & $71$ & $27$ & $14$ &  & $200$ &  & $102$ & $27$ & $16$ \\ 
$210$ &  & $75$ & $28$ & $15$ &  & $210$ &  & $107$ & $28$ & $17$ \\ 
$220$ &  & $80$ & $29$ & $16$ &  & $220$ &  & $109$ & $29$ & $18$ \\ 
$230$ &  & $92$ & $30$ & $17$ &  & $230$ &  & $110$ & $30$ & $18$ \\ 
$240$ &  & $103$ & $31$ & $18$ &  & $240$ &  & $112$ & $31$ & $19$ \\ 
$250$ &  & $108$ & $32$ & $18$ &  & $250$ &  & $112$ & $33$ & $20$ \\ 
$260$ &  & $111$ & $34$ & $19$ &  & $260$ &  & $111$ & $34$ & $21$ \\ 
$270$ &  & $112$ & $35$ & $20$ &  & $270$ &  & $110$ & $35$ & $22$ \\ 
$280$ &  & $112$ & $36$ & $21$ &  & $280$ &  & $108$ & $36$ & $23$ \\ 
$290$ &  & $112$ & $37$ & $22$ &  & $290$ &  & $106$ & $37$ & $24$ \\ 
$300$ &  & $111$ & $38$ & $23$ &  & $300$ &  & $103$ & $38$ & $25$ \\ 
$310$ &  & $108$ & $39$ & $23$ &  & $310$ &  & $99$ & $39$ & $26$ \\ 
$320$ &  & $105$ & $40$ & $24$ &  & $320$ &  & $94$ & $40$ & $26$ \\ 
$330$ &  & $101$ & $41$ & $25$ &  & $330$ &  & $87$ & $41$ & $26$ \\ 
$340$ &  & $95$ & $42$ & $26$ &  & $340$ &  & $77$ & $43$ & $26$ \\ 
$350$ &  & $87$ & $43$ & $26$ &  & $350$ &  & $70$ & $43$ & $23$ \\ 
$360$ &  & $82$ & $44$ & $25$ &  & $360$ &  & $74$ & $44$ & $20$ \\ 
$370$ &  & $82$ & $46$ & $25$ &  & $370$ &  & $78$ & $46$ & $19$ \\ 
$380$ &  & $81$ & $47$ & $24$ &  & $380$ &  & $81$ & $47$ & $17$ \\ 
$390$ &  & $81$ & $48$ & $24$ &  & $390$ &  & $83$ & $48$ & $15$ \\ 
\bottomrule
\end{tabular}
\end{center}
\caption{Values of the scales $w_{t,b,i}^{gg+qg}$ as a function of the scalar and pseudoscalar Higgs mass.}
\label{tab:wtwbwiappendix1}
\end{table}
\begin{table}[!h]
\centering
\begin{center}
\begin{tabular}{@{}cccccc|ccccc}
\toprule
\multicolumn{11}{c}{\bf Scalar and pseudoscalar collinear deviation scale $w$ (GeV)}\\
\midrule
$\mh$ (GeV) & & $w^{gg+qg}_t$ & $w^{gg+qg}_b$ & $w^{gg+qg}_i$ & & $m_A$ (GeV) & & $w^{gg+qg}_t$ & $w^{gg+qg}_b$ & $w^{gg+qg}_i$ \\
\cmidrule{1-1} \cmidrule{3-5} \cmidrule{7-7} \cmidrule{9-11} 
$400$ &  & $81$ & $49$ & $23$ &  & $400$ &  & $86$ & $49$ & $14$ \\ 
$410$ &  & $82$ & $50$ & $23$ &  & $410$ &  & $88$ & $50$ & $12$ \\ 
$420$ &  & $83$ & $51$ & $23$ &  & $420$ &  & $91$ & $51$ & $11$ \\ 
$430$ &  & $85$ & $52$ & $22$ &  & $430$ &  & $93$ & $52$ & $9$ \\ 
$440$ &  & $86$ & $53$ & $21$ &  & $440$ &  & $95$ & $53$ & $6$ \\ 
$450$ &  & $87$ & $53$ & $21$ &  & $450$ &  & $98$ & $53$ & $2$ \\ 
$460$ &  & $89$ & $54$ & $20$ &  & $460$ &  & $100$ & $54$ & $6$ \\ 
$470$ &  & $91$ & $55$ & $19$ &  & $470$ &  & $102$ & $55$ & $8$ \\ 
$480$ &  & $92$ & $56$ & $19$ &  & $480$ &  & $105$ & $56$ & $10$ \\ 
$490$ &  & $94$ & $57$ & $18$ &  & $490$ &  & $107$ & $57$ & $12$ \\
$500$ &  & $96$ & $58$ & $17$ &  & $500$ &  & $109$ & $58$ & $14$ \\ 
$510$ &  & $97$ & $59$ & $16$ &  & $510$ &  & $112$ & $59$ & $16$ \\ 
$520$ &  & $99$ & $60$ & $15$ &  & $520$ &  & $114$ & $60$ & $18$ \\ 
$530$ &  & $101$ & $61$ & $14$ &  & $530$ &  & $116$ & $61$ & $20$ \\ 
$540$ &  & $102$ & $62$ & $12$ &  & $540$ &  & $118$ & $62$ & $21$ \\ 
$550$ &  & $104$ & $63$ & $11$ &  & $550$ &  & $120$ & $63$ & $23$ \\ 
$560$ &  & $106$ & $64$ & $10$ &  & $560$ &  & $122$ & $64$ & $25$ \\ 
$570$ &  & $107$ & $65$ & $8$ &  & $570$ &  & $125$ & $65$ & $27$ \\ 
$580$ &  & $109$ & $66$ & $5$ &  & $580$ &  & $127$ & $66$ & $29$ \\ 
$590$ &  & $111$ & $67$ & $2$ &  & $590$ &  & $129$ & $67$ & $38$ \\ 
$600$ &  & $113$ & $68$ & $6$ &  & $600$ &  & $132$ & $68$ & $39$ \\ 
$610$ &  & $115$ & $69$ & $8$ &  & $610$ &  & $134$ & $69$ & $40$ \\ 
$620$ &  & $117$ & $70$ & $10$ &  & $620$ &  & $136$ & $70$ & $41$ \\ 
$630$ &  & $119$ & $71$ & $12$ &  & $630$ &  & $138$ & $71$ & $43$ \\ 
$640$ &  & $121$ & $72$ & $14$ &  & $640$ &  & $140$ & $72$ & $44$ \\ 
$650$ &  & $127$ & $73$ & $16$ &  & $650$ &  & $143$ & $73$ & $46$ \\ 
$660$ &  & $129$ & $74$ & $17$ &  & $660$ &  & $145$ & $74$ & $48$ \\ 
$670$ &  & $131$ & $75$ & $19$ &  & $670$ &  & $153$ & $75$ & $51$ \\ 
$680$ &  & $133$ & $76$ & $21$ &  & $680$ &  & $155$ & $76$ & $54$ \\ 
$690$ &  & $135$ & $77$ & $22$ &  & $690$ &  & $157$ & $77$ & $57$ \\ 
$700$ &  & $137$ & $78$ & $24$ &  & $700$ &  & $160$ & $77$ & $60$ \\ 
$710$ &  & $139$ & $79$ & $25$ &  & $710$ &  & $162$ & $78$ & $64$ \\ 
$720$ &  & $141$ & $79$ & $27$ &  & $720$ &  & $165$ & $79$ & $68$ \\ 
$730$ &  & $143$ & $80$ & $28$ &  & $730$ &  & $167$ & $80$ & $74$ \\ 
$740$ &  & $146$ & $81$ & $30$ &  & $740$ &  & $170$ & $81$ & $81$ \\ 
$750$ &  & $148$ & $82$ & $32$ &  & $750$ &  & $172$ & $82$ & $91$ \\ 
$760$ &  & $150$ & $83$ & $33$ &  & $760$ &  & $174$ & $83$ & $235$ \\ 
$770$ &  & $152$ & $84$ & $35$ &  & $770$ &  & $177$ & $84$ & $234$ \\ 
$780$ &  & $154$ & $85$ & $44$ &  & $780$ &  & $179$ & $84$ & $233$ \\ 
$790$ &  & $156$ & $86$ & $45$ &  & $790$ &  & $181$ & $85$ & $232$ \\ 
$800$ &  & $158$ & $87$ & $46$ &  & $800$ &  & $184$ & $86$ & $239$ \\ 
\bottomrule
\end{tabular}
\end{center}
\caption{Values of the scales $w_{t,b,i}^{gg+qg}$ as a function of the scalar and pseudoscalar Higgs mass.}
\label{tab:wtwbwiappendix2}
\end{table}
\clearpage

\bibliographystyle{unsrt}
\bibliography{pth}

\begin{thebibliography}{100}

\bibitem{Grazzini:2013mca}
Massimiliano Grazzini and Hayk Sargsyan.
\newblock {Heavy-quark mass effects in Higgs boson production at the LHC}.
\newblock {\em JHEP}, 1309:129, 2013.

\bibitem{Aad:2012tfa}
Georges Aad et~al.
\newblock {Observation of a new particle in the search for the Standard Model
  Higgs boson with the ATLAS detector at the LHC}.
\newblock {\em Phys.Lett.}, B716:1--29, 2012.

\bibitem{Chatrchyan:2012ufa}
Serguei Chatrchyan et~al.
\newblock {Observation of a new boson at a mass of 125 GeV with the CMS
  experiment at the LHC}.
\newblock {\em Phys.Lett.}, B716:30--61, 2012.

\bibitem{Dittmaier:2011ti}
S.~Dittmaier et~al.
\newblock {Handbook of LHC Higgs Cross Sections: 1. Inclusive Observables}.
\newblock 2011.

\bibitem{Dittmaier:2012vm}
S.~Dittmaier, S.~Dittmaier, C.~Mariotti, G.~Passarino, R.~Tanaka, et~al.
\newblock {Handbook of LHC Higgs Cross Sections: 2. Differential
  Distributions}.
\newblock 2012.

\bibitem{Heinemeyer:2013tqa}
S~Heinemeyer et~al.
\newblock {Handbook of LHC Higgs Cross Sections: 3. Higgs Properties}.
\newblock 2013.

\bibitem{Dawson:1990zj}
S.~Dawson.
\newblock {Radiative corrections to Higgs boson production}.
\newblock {\em Nucl.Phys.}, B359:283--300, 1991.

\bibitem{Djouadi:1991tka}
A.~Djouadi, M.~Spira, and P.M. Zerwas.
\newblock {Production of Higgs bosons in proton colliders: QCD corrections}.
\newblock {\em Phys.Lett.}, B264:440--446, 1991.

\bibitem{Spira:1995rr}
M.~Spira, A.~Djouadi, D.~Graudenz, and P.M. Zerwas.
\newblock {Higgs boson production at the LHC}.
\newblock {\em Nucl.Phys.}, B453:17--82, 1995.

\bibitem{Harlander:2000mg}
Robert~V. Harlander.
\newblock {Virtual corrections to g g $\to$ H to two loops in the heavy top
  limit}.
\newblock {\em Phys.Lett.}, B492:74--80, 2000.

\bibitem{Harlander:2001is}
Robert~V. Harlander and William~B. Kilgore.
\newblock {Soft and virtual corrections to proton proton $\to$ H + x at NNLO}.
\newblock {\em Phys.Rev.}, D64:013015, 2001.

\bibitem{Catani:2001ic}
Stefano Catani, Daniel de~Florian, and Massimiliano Grazzini.
\newblock {Higgs production in hadron collisions: Soft and virtual QCD
  corrections at NNLO}.
\newblock {\em JHEP}, 0105:025, 2001.

\bibitem{Harlander:2002wh}
Robert~V. Harlander and William~B. Kilgore.
\newblock {Next-to-next-to-leading order Higgs production at hadron colliders}.
\newblock {\em Phys.Rev.Lett.}, 88:201801, 2002.

\bibitem{Anastasiou:2002yz}
Charalampos Anastasiou and Kirill Melnikov.
\newblock {Higgs boson production at hadron colliders in NNLO QCD}.
\newblock {\em Nucl.Phys.}, B646:220--256, 2002.

\bibitem{Ravindran:2003um}
V.~Ravindran, J.~Smith, and W.~L. van Neerven.
\newblock {NNLO corrections to the total cross-section for Higgs boson
  production in hadron hadron collisions}.
\newblock {\em Nucl.Phys.}, B665:325--366, 2003.

\bibitem{Moch:2005ky}
S.~Moch and A.~Vogt.
\newblock {Higher-order soft corrections to lepton pair and Higgs boson
  production}.
\newblock {\em Phys.Lett.}, B631:48--57, 2005.

\bibitem{Ravindran:2006cg}
V.~Ravindran.
\newblock {Higher-order threshold effects to inclusive processes in QCD}.
\newblock {\em Nucl.Phys.}, B752:173--196, 2006.

\bibitem{Ball:2013bra}
Richard~D. Ball, Marco Bonvini, Stefano Forte, Simone Marzani, and Giovanni
  Ridolfi.
\newblock {Higgs production in gluon fusion beyond NNLO}.
\newblock {\em Nucl.Phys.}, B874:746--772, 2013.

\bibitem{Buehler:2013fha}
Stephan Buehler and Achilleas Lazopoulos.
\newblock {Scale dependence and collinear subtraction terms for Higgs
  production in gluon fusion at N3LO}.
\newblock {\em JHEP}, 1310:096, 2013.

\bibitem{Anastasiou:2014vaa}
Charalampos Anastasiou, Claude Duhr, Falko Dulat, Elisabetta Furlan, Thomas
  Gehrmann, et~al.
\newblock {Higgs boson gluon-fusion production at threshold in N3LO QCD}.
\newblock 2014.

\bibitem{Anastasiou:2015ema}
Charalampos Anastasiou, Claude Duhr, Falko Dulat, Franz Herzog, and Bernhard
  Mistlberger.
\newblock {Higgs boson gluon-fusion production in N3LO QCD}.
\newblock 2015.

\bibitem{Harlander:2005rq}
Robert Harlander and Philipp Kant.
\newblock {Higgs production and decay: Analytic results at next-to-leading
  order QCD}.
\newblock {\em JHEP}, 0512:015, 2005.

\bibitem{Anastasiou:2006hc}
Charalampos Anastasiou, Stefan Beerli, Stefan Bucherer, Alejandro Daleo, and
  Zoltan Kunszt.
\newblock {Two-loop amplitudes and master integrals for the production of a
  Higgs boson via a massive quark and a scalar-quark loop}.
\newblock {\em JHEP}, 0701:082, 2007.

\bibitem{Aglietti:2006tp}
U.~Aglietti, R.~Bonciani, G.~Degrassi, and A.~Vicini.
\newblock {Analytic Results for Virtual QCD Corrections to Higgs Production and
  Decay}.
\newblock {\em JHEP}, 0701:021, 2007.

\bibitem{Bonciani:2007ex}
R.~Bonciani, Giuseppe Degrassi, and A.~Vicini.
\newblock {Scalar particle contribution to Higgs production via gluon fusion at
  NLO}.
\newblock {\em JHEP}, 0711:095, 2007.

\bibitem{Marzani:2008az}
Simone Marzani, Richard~D. Ball, Vittorio Del~Duca, Stefano Forte, and
  Alessandro Vicini.
\newblock {Higgs production via gluon-gluon fusion with finite top mass beyond
  next-to-leading order}.
\newblock {\em Nucl.Phys.}, B800:127--145, 2008.

\bibitem{Harlander:2009bw}
Robert~V. Harlander and Kemal~J. Ozeren.
\newblock {Top mass effects in Higgs production at next-to-next-to-leading
  order QCD: Virtual corrections}.
\newblock {\em Phys.Lett.}, B679:467--472, 2009.

\bibitem{Harlander:2009mq}
Robert~V. Harlander and Kemal~J. Ozeren.
\newblock {Finite top mass effects for hadronic Higgs production at
  next-to-next-to-leading order}.
\newblock {\em JHEP}, 0911:088, 2009.

\bibitem{Pak:2009bx}
Alexey Pak, Mikhail Rogal, and Matthias Steinhauser.
\newblock {Virtual three-loop corrections to Higgs boson production in gluon
  fusion for finite top quark mass}.
\newblock {\em Phys.Lett.}, B679:473--477, 2009.

\bibitem{Pak:2009dg}
Alexey Pak, Mikhail Rogal, and Matthias Steinhauser.
\newblock {Finite top quark mass effects in NNLO Higgs boson production at
  LHC}.
\newblock {\em JHEP}, 1002:025, 2010.

\bibitem{Harlander:2009my}
Robert~V. Harlander, Hendrik Mantler, Simone Marzani, and Kemal~J. Ozeren.
\newblock {Higgs production in gluon fusion at next-to-next-to-leading order
  QCD for finite top mass}.
\newblock {\em Eur.Phys.J.}, C66:359--372, 2010.

\bibitem{Kramer:1996iq}
Michael Kramer, Eric Laenen, and Michael Spira.
\newblock {Soft gluon radiation in Higgs boson production at the LHC}.
\newblock {\em Nucl.Phys.}, B511:523--549, 1998.

\bibitem{Catani:2003zt}
Stefano Catani, Daniel de~Florian, Massimiliano Grazzini, and Paolo Nason.
\newblock {Soft gluon resummation for Higgs boson production at hadron
  colliders}.
\newblock {\em JHEP}, 0307:028, 2003.

\bibitem{Idilbi:2005ni}
Ahmad Idilbi, Xiang-dong Ji, Jian-Ping Ma, and Feng Yuan.
\newblock {Threshold resummation for Higgs production in effective field
  theory}.
\newblock {\em Phys.Rev.}, D73:077501, 2006.

\bibitem{Idilbi:2006dg}
Ahmad Idilbi, Xiang-dong Ji, and Feng Yuan.
\newblock {Resummation of threshold logarithms in effective field theory for
  DIS, Drell-Yan and Higgs production}.
\newblock {\em Nucl.Phys.}, B753:42--68, 2006.

\bibitem{Ahrens:2008nc}
Valentin Ahrens, Thomas Becher, Matthias Neubert, and Li~Lin Yang.
\newblock {Renormalization-Group Improved Prediction for Higgs Production at
  Hadron Colliders}.
\newblock {\em Eur.Phys.J.}, C62:333--353, 2009.

\bibitem{Djouadi:1994ge}
A.~Djouadi and P.~Gambino.
\newblock {Leading electroweak correction to Higgs boson production at proton
  colliders}.
\newblock {\em Phys.Rev.Lett.}, 73:2528--2531, 1994.

\bibitem{Djouadi:1997rj}
A.~Djouadi, P.~Gambino, and Bernd~A. Kniehl.
\newblock {Two loop electroweak heavy fermion corrections to Higgs boson
  production and decay}.
\newblock {\em Nucl.Phys.}, B523:17--39, 1998.

\bibitem{Aglietti:2004nj}
U.~Aglietti, R.~Bonciani, G.~Degrassi, and A.~Vicini.
\newblock {Two loop light fermion contribution to Higgs production and decays}.
\newblock {\em Phys.Lett.}, B595:432--441, 2004.

\bibitem{Aglietti:2004ki}
U.~Aglietti, R.~Bonciani, G.~Degrassi, and A.~Vicini.
\newblock {Master integrals for the two-loop light fermion contributions to gg
  $\to$ H and H $\to$ gamma gamma}.
\newblock {\em Phys.Lett.}, B600:57--64, 2004.

\bibitem{Degrassi:2004mx}
Giuseppe Degrassi and Fabio Maltoni.
\newblock {Two-loop electroweak corrections to Higgs production at hadron
  colliders}.
\newblock {\em Phys.Lett.}, B600:255--260, 2004.

\bibitem{Actis:2008ug}
Stefano Actis, Giampiero Passarino, Christian Sturm, and Sandro Uccirati.
\newblock {NLO Electroweak Corrections to Higgs Boson Production at Hadron
  Colliders}.
\newblock {\em Phys.Lett.}, B670:12--17, 2008.

\bibitem{Actis:2008ts}
Stefano Actis, Giampiero Passarino, Christian Sturm, and Sandro Uccirati.
\newblock {NNLO Computational Techniques: The Cases H $\to$ gamma gamma and H
  $\to$ g g}.
\newblock {\em Nucl.Phys.}, B811:182--273, 2009.

\bibitem{Bonciani:2010ms}
R.~Bonciani, G.~Degrassi, and A.~Vicini.
\newblock {On the Generalized Harmonic Polylogarithms of One Complex Variable}.
\newblock {\em Comput.Phys.Commun.}, 182:1253--1264, 2011.

\bibitem{Anastasiou:2008tj}
Charalampos Anastasiou, Radja Boughezal, and Frank Petriello.
\newblock {Mixed QCD-electroweak corrections to Higgs boson production in gluon
  fusion}.
\newblock {\em JHEP}, 0904:003, 2009.

\bibitem{Demartin:2010er}
Federico Demartin, Stefano Forte, Elisa Mariani, Juan Rojo, and Alessandro
  Vicini.
\newblock {The impact of PDF and alphas uncertainties on Higgs Production in
  gluon fusion at hadron colliders}.
\newblock {\em Phys.Rev.}, D82:014002, 2010.

\bibitem{Ellis:1987xu}
R.~Keith Ellis, I.~Hinchliffe, M.~Soldate, and J.J. van~der Bij.
\newblock {Higgs Decay to tau+ tau-: A Possible Signature of Intermediate Mass
  Higgs Bosons at the SSC}.
\newblock {\em Nucl.Phys.}, B297:221, 1988.

\bibitem{Baur:1989cm}
U.~Baur and E.W.~Nigel Glover.
\newblock {Higgs Boson Production at Large Transverse Momentum in Hadronic
  Collisions}.
\newblock {\em Nucl.Phys.}, B339:38--66, 1990.

\bibitem{Keung:2009bs}
Wai-Yee Keung and Frank~J. Petriello.
\newblock {Electroweak and finite quark-mass effects on the Higgs boson
  transverse momentum distribution}.
\newblock {\em Phys.Rev.}, D80:013007, 2009.

\bibitem{Brein:2010xj}
Oliver Brein.
\newblock {Electroweak and Bottom Quark Contributions to Higgs Boson plus Jet
  Production}.
\newblock {\em Phys.Rev.}, D81:093006, 2010.

\bibitem{deFlorian:1999zd}
D.~de~Florian, M.~Grazzini, and Z.~Kunszt.
\newblock {Higgs production with large transverse momentum in hadronic
  collisions at next-to-leading order}.
\newblock {\em Phys.Rev.Lett.}, 82:5209--5212, 1999.

\bibitem{Ravindran:2002dc}
V.~Ravindran, J.~Smith, and W.L. Van~Neerven.
\newblock {Next-to-leading order QCD corrections to differential distributions
  of Higgs boson production in hadron hadron collisions}.
\newblock {\em Nucl.Phys.}, B634:247--290, 2002.

\bibitem{Glosser:2002gm}
Christopher~J. Glosser and Carl~R. Schmidt.
\newblock {Next-to-leading corrections to the Higgs boson transverse momentum
  spectrum in gluon fusion}.
\newblock {\em JHEP}, 0212:016, 2002.

\bibitem{Harlander:2012hf}
Robert~V. Harlander, Tobias Neumann, Kemal~J. Ozeren, and Marius Wiesemann.
\newblock {Top-mass effects in differential Higgs production through gluon
  fusion at order $\alpha_s^4$}.
\newblock {\em JHEP}, 1208:139, 2012.

\bibitem{Boughezal:2013uia}
Radja Boughezal, Fabrizio Caola, Kirill Melnikov, Frank Petriello, and Markus
  Schulze.
\newblock {Higgs boson production in association with a jet at
  next-to-next-to-leading order in perturbative QCD}.
\newblock {\em JHEP}, 1306:072, 2013.

\bibitem{Bozzi:2003jy}
G.~Bozzi, S.~Catani, D.~de~Florian, and M.~Grazzini.
\newblock {The q(T) spectrum of the Higgs boson at the LHC in QCD perturbation
  theory}.
\newblock {\em Phys.Lett.}, B564:65--72, 2003.

\bibitem{Bozzi:2005wk}
Giuseppe Bozzi, Stefano Catani, Daniel de~Florian, and Massimiliano Grazzini.
\newblock {Transverse-momentum resummation and the spectrum of the Higgs boson
  at the LHC}.
\newblock {\em Nucl.Phys.}, B737:73--120, 2006.

\bibitem{Bozzi:2007pn}
Giuseppe Bozzi, Stefano Catani, Daniel de~Florian, and Massimiliano Grazzini.
\newblock {Higgs boson production at the LHC: Transverse-momentum resummation
  and rapidity dependence}.
\newblock {\em Nucl.Phys.}, B791:1--19, 2008.

\bibitem{deFlorian:2012mx}
D.~de~Florian, G.~Ferrera, M.~Grazzini, and D.~Tommasini.
\newblock {Higgs boson production at the LHC: transverse momentum resummation
  effects in the $H \to \gamma\gamma$, $H \to WW \to l\nu l \nu$ and $H \to ZZ
  \to 4l$ decay modes}.
\newblock {\em JHEP}, 1206:132, 2012.

\bibitem{Chiu:2012ir}
Jui-Yu Chiu, Ambar Jain, Duff Neill, and Ira~Z. Rothstein.
\newblock {A Formalism for the Systematic Treatment of Rapidity Logarithms in
  Quantum Field Theory}.
\newblock {\em JHEP}, 1205:084, 2012.

\bibitem{Becher:2012yn}
Thomas Becher, Matthias Neubert, and Daniel Wilhelm.
\newblock {Higgs-Boson Production at Small Transverse Momentum}.
\newblock {\em JHEP}, 1305:110, 2013.

\bibitem{Neill:2015roa}
Duff Neill, Ira~Z. Rothstein, and Varun Vaidya.
\newblock {The Higgs Transverse Momentum Distribution at NNLL and its
  Theoretical Errors}.
\newblock 2015.

\bibitem{Frixione:2002ik}
Stefano Frixione and Bryan~R. Webber.
\newblock {Matching NLO QCD computations and parton shower simulations}.
\newblock {\em JHEP}, 0206:029, 2002.

\bibitem{Alioli:2008tz}
Simone Alioli, Paolo Nason, Carlo Oleari, and Emanuele Re.
\newblock {NLO Higgs boson production via gluon fusion matched with shower in
  POWHEG}.
\newblock {\em JHEP}, 0904:002, 2009.

\bibitem{Hamilton:2013fea}
Keith Hamilton, Paolo Nason, Emanuele Re, and Giulia Zanderighi.
\newblock {NNLOPS simulation of Higgs boson production}.
\newblock {\em JHEP}, 1310:222, 2013.

\bibitem{Hoche:2014dla}
Stefan Höche, Ye~Li, and Stefan Prestel.
\newblock {Higgs-boson production through gluon fusion at NNLO QCD with parton
  showers}.
\newblock {\em Phys.Rev.}, D90(5):054011, 2014.

\bibitem{Bagnaschi:2011tu}
E.~Bagnaschi, G.~Degrassi, P.~Slavich, and A.~Vicini.
\newblock {Higgs production via gluon fusion in the POWHEG approach in the SM
  and in the MSSM}.
\newblock {\em JHEP}, 1202:088, 2012.

\bibitem{frixione-masses}
Stefano Frixione.
\newblock New developments in nlo mc.
\newblock Presented at the 7th Workshop of LHC Higgs Cross Section Working
  Group at CERN ,
  https://indico.cern.ch/event/209605/session/5/contribution/18/material/slides/0.pdf,
  CERN, December 6th 2012.

\bibitem{Mantler:2012bj}
Hendrik Mantler and Marius Wiesemann.
\newblock {Top- and bottom-mass effects in hadronic Higgs production at small
  transverse momenta through LO+NLL}.
\newblock {\em Eur.Phys.J.}, C73:2467, 2013.

\bibitem{Hamilton:2015nsa}
Keith Hamilton, Paolo Nason, and Giulia Zanderighi.
\newblock {Finite quark-mass effects in the NNLOPS POWHEG+MiNLO Higgs
  generator}.
\newblock 2015.

\bibitem{Banfi:2013eda}
Andrea Banfi, Pier~Francesco Monni, and Giulia Zanderighi.
\newblock {Quark masses in Higgs production with a jet veto}.
\newblock {\em JHEP}, 1401:097, 2014.

\bibitem{Neumann:2014nha}
Tobias Neumann and Marius Wiesemann.
\newblock {Finite top-mass effects in gluon-induced Higgs production with a
  jet-veto at NNLO}.
\newblock {\em JHEP}, 1411:150, 2014.

\bibitem{Ligeti:2008ac}
Zoltan Ligeti, Iain~W. Stewart, and Frank~J. Tackmann.
\newblock {Treating the b quark distribution function with reliable
  uncertainties}.
\newblock {\em Phys.Rev.}, D78:114014, 2008.

\bibitem{Abbate:2010xh}
Riccardo Abbate, Michael Fickinger, Andre~H. Hoang, Vicent Mateu, and Iain~W.
  Stewart.
\newblock {Thrust at N$^3$LL with Power Corrections and a Precision Global Fit
  for alphas(mZ)}.
\newblock {\em Phys.Rev.}, D83:074021, 2011.

\bibitem{Berger:2010xi}
Carola~F. Berger, Claudio Marcantonini, Iain~W. Stewart, Frank~J. Tackmann, and
  Wouter~J. Waalewijn.
\newblock {Higgs Production with a Central Jet Veto at NNLL+NNLO}.
\newblock {\em JHEP}, 1104:092, 2011.

\bibitem{cerntalk}
Alessandro Vicini.
\newblock Higgs transverse momentum distribution in shower montecarlo codes for
  pp$\to$h+x.
\newblock Presented at the ATLAS workshop (N)NLO Monte Carlo generators for LHC
  Run 2, http://wwwteor.mi.infn.it/~vicini/slides/Higgs\_pt\_16dec13.pdf, CERN,
  December 16th 2013.

\bibitem{Harlander:2014uea}
Robert~V. Harlander, Hendrik Mantler, and Marius Wiesemann.
\newblock {Transverse momentum resummation for Higgs production via gluon
  fusion in the MSSM}.
\newblock {\em JHEP}, 1411:116, 2014.

\bibitem{Mantler:2015vba}
Hendrik Mantler and Marius Wiesemann.
\newblock {Hadronic Higgs production through NLO $+$ PS in the SM, the 2HDM and
  the MSSM}.
\newblock {\em Eur. Phys. J.}, C75(6):257, 2015.

\bibitem{Bagnaschi:2015bop}
Emanuele Bagnaschi, Robert~V. Harlander, Hendrik Mantler, Alessandro Vicini,
  and Marius Wiesemann.
\newblock {Resummation ambiguities in the Higgs transverse-momentum spectrum in
  the Standard Model and beyond}.
\newblock 2015.

\bibitem{Bagnaschi:2014zla}
E.~Bagnaschi, R.V. Harlander, S.~Liebler, H.~Mantler, P.~Slavich, et~al.
\newblock {Towards precise predictions for Higgs-boson production in the MSSM}.
\newblock {\em JHEP}, 1406:167, 2014.

\bibitem{Dokshitzer:1978hw}
Yuri~L. Dokshitzer, Dmitri Diakonov, and S.I. Troian.
\newblock {Hard Processes in Quantum Chromodynamics}.
\newblock {\em Phys.Rept.}, 58:269--395, 1980.

\bibitem{Parisi:1979se}
G.~Parisi and R.~Petronzio.
\newblock {Small Transverse Momentum Distributions in Hard Processes}.
\newblock {\em Nucl.Phys.}, B154:427, 1979.

\bibitem{Curci:1979bg}
G.~Curci, Mario Greco, and Y.~Srivastava.
\newblock {{QCD} Jets From Coherent States}.
\newblock {\em Nucl.Phys.}, B159:451, 1979.

\bibitem{Collins:1981uk}
John~C. Collins and Davison~E. Soper.
\newblock {Back-To-Back Jets in QCD}.
\newblock {\em Nucl.Phys.}, B193:381, 1981.

\bibitem{Collins:1981va}
John~C. Collins and Davison~E. Soper.
\newblock {Back-To-Back Jets: Fourier Transform from B to K-Transverse}.
\newblock {\em Nucl.Phys.}, B197:446, 1982.

\bibitem{Kodaira:1981nh}
Jiro Kodaira and Luca Trentadue.
\newblock {Summing Soft Emission in QCD}.
\newblock {\em Phys.Lett.}, B112:66, 1982.

\bibitem{Kodaira:1982az}
Jiro Kodaira and Luca Trentadue.
\newblock {Single Logarithm Effects in electron-Positron Annihilation}.
\newblock {\em Phys.Lett.}, B123:335, 1983.

\bibitem{Altarelli:1984pt}
Guido Altarelli, R.~Keith Ellis, Mario Greco, and G.~Martinelli.
\newblock {Vector Boson Production at Colliders: A Theoretical Reappraisal}.
\newblock {\em Nucl.Phys.}, B246:12, 1984.

\bibitem{Collins:1984kg}
John~C. Collins, Davison~E. Soper, and George~F. Sterman.
\newblock {Transverse Momentum Distribution in Drell-Yan Pair and W and Z Boson
  Production}.
\newblock {\em Nucl.Phys.}, B250:199, 1985.

\bibitem{Catani:2000vq}
Stefano Catani, Daniel de~Florian, and Massimiliano Grazzini.
\newblock {Universality of nonleading logarithmic contributions in transverse
  momentum distributions}.
\newblock {\em Nucl.Phys.}, B596:299--312, 2001.

\bibitem{Frixione:2007vw}
Stefano Frixione, Paolo Nason, and Carlo Oleari.
\newblock {Matching NLO QCD computations with Parton Shower simulations: the
  POWHEG method}.
\newblock {\em JHEP}, 0711:070, 2007.

\bibitem{Alioli:2010xd}
Simone Alioli, Paolo Nason, Carlo Oleari, and Emanuele Re.
\newblock {A general framework for implementing NLO calculations in shower
  Monte Carlo programs: the POWHEG BOX}.
\newblock {\em JHEP}, 1006:043, 2010.

\bibitem{Degrassi:2011vq}
G.~Degrassi, S.~Di~Vita, and P.~Slavich.
\newblock {NLO QCD corrections to pseudoscalar Higgs production in the MSSM}.
\newblock {\em JHEP}, 1108:128, 2011.

\bibitem{Harlander:2012pb}
Robert~V. Harlander, Stefan Liebler, and Hendrik Mantler.
\newblock {SusHi: A program for the calculation of Higgs production in gluon
  fusion and bottom-quark annihilation in the Standard Model and the MSSM}.
\newblock {\em Computer Physics Communications}, 184:1605--1617, 2013.

\bibitem{deFlorian:2011xf}
Daniel de~Florian, Giancarlo Ferrera, Massimiliano Grazzini, and Damiano
  Tommasini.
\newblock {Transverse-momentum resummation: Higgs boson production at the
  Tevatron and the LHC}.
\newblock {\em JHEP}, 1111:064, 2011.

\bibitem{Martin:2009iq}
A.D. Martin, W.J. Stirling, R.S. Thorne, and G.~Watt.
\newblock {Parton distributions for the LHC}.
\newblock {\em Eur.Phys.J.}, C63:189--285, 2009.

\bibitem{Sjostrand:2007gs}
Torbjorn Sjostrand, Stephen Mrenna, and Peter~Z. Skands.
\newblock {A Brief Introduction to PYTHIA 8.1}.
\newblock {\em Comput.Phys.Commun.}, 178:852--867, 2008.

\bibitem{Sjostrand:2006za}
Torbjorn Sjostrand, Stephen Mrenna, and Peter~Z. Skands.
\newblock {PYTHIA 6.4 Physics and Manual}.
\newblock {\em JHEP}, 0605:026, 2006.

\bibitem{Carena:2013qia}
M.~Carena, S.~Heinemeyer, O.~Stål, C.E.M. Wagner, and G.~Weiglein.
\newblock {MSSM Higgs Boson Searches at the LHC: Benchmark Scenarios after the
  Discovery of a Higgs-like Particle}.
\newblock {\em Eur.Phys.J.}, C73:2552, 2013.

\bibitem{Bechtle:2008jh}
Philip Bechtle, Oliver Brein, Sven Heinemeyer, Georg Weiglein, and Karina~E.
  Williams.
\newblock {HiggsBounds: Confronting Arbitrary Higgs Sectors with Exclusion
  Bounds from LEP and the Tevatron}.
\newblock {\em Comput.Phys.Commun.}, 181:138--167, 2010.

\bibitem{Bechtle:2011sb}
Philip Bechtle, Oliver Brein, Sven Heinemeyer, Georg Weiglein, and Karina~E.
  Williams.
\newblock {HiggsBounds 2.0.0: Confronting Neutral and Charged Higgs Sector
  Predictions with Exclusion Bounds from LEP and the Tevatron}.
\newblock {\em Comput.Phys.Commun.}, 182:2605--2631, 2011.

\bibitem{Bechtle:2013gu}
Philip Bechtle, Oliver Brein, Sven Heinemeyer, Oscar Stal, Tim Stefaniak,
  et~al.
\newblock {Recent Developments in HiggsBounds and a Preview of HiggsSignals}.
\newblock {\em PoS}, CHARGED2012:024, 2012.

\bibitem{Bechtle:2013wla}
Philip Bechtle, Oliver Brein, Sven Heinemeyer, Oscar Stål, Tim Stefaniak,
  et~al.
\newblock {$\mathsf{HiggsBounds}-4$: Improved Tests of Extended Higgs Sectors
  against Exclusion Bounds from LEP, the Tevatron and the LHC}.
\newblock {\em Eur.Phys.J.}, C74(3):2693, 2014.

\end{thebibliography}
\end{document}